\documentclass[final, 1p, times]{elsarticle}
\usepackage{float}
\usepackage[pagewise]{lineno}
\usepackage[dvipsnames]{xcolor}
\usepackage{framed}
\usepackage{multicol}
\usepackage{amssymb}
\usepackage{todonotes}     
\usepackage[utf8]{inputenc}
\usepackage[ruled, vlined]{algorithm2e}
\usepackage{graphicx}
\usepackage{subcaption}
\usepackage{tabularx}
\usepackage{amsmath}
\usepackage{multirow}
\usepackage{setspace}
\usepackage{tabu}

\DeclareMathAlphabet{\mathcal}{OMS}{cmsy}{m}{n}
\DeclareMathAlphabet\mathbfcal{OMS}{cmsy}{b}{n}
\newcommand{\mat}[1]{\boldsymbol{#1}}

\usepackage{siunitx}
\usepackage[colorlinks, allcolors=blue]{hyperref}
\usepackage[noabbrev]{cleveref}

\usepackage[percent]{overpic}

\usepackage[normalem]{ulem}
\usepackage{color}

\begin{document}
\begin{frontmatter}
\title{Recurrent neural networks and Koopman-based frameworks for temporal predictions in a low-order model of turbulence}

\author[1]{Hamidreza Eivazi}
\ead{hamid.eivazi@ut.ac.ir}

\author[2,3]{Luca Guastoni}

\author[2,3]{Philipp Schlatter}

\author[4,3]{Hossein Azizpour}

\author[2,3]{Ricardo Vinuesa\corref{cor}}
\ead{rvinuesa@mech.kth.se}

\address[1]{Faculty of New Sciences and Technologies, University of Tehran, Tehran, Iran}
\address[2]{SimEx/FLOW, Engineering Mechanics, KTH Royal Institute of Technology, \\ SE-100 44 Stockholm, Sweden}
\address[3]{Swedish e-Science Research Centre (SeRC), Stockholm, Sweden}
\address[4]{Division of Robotics, Perception, and Learning, School of EECS, KTH Royal Institute of Technology, Stockholm, Sweden}


\cortext[cor]{Corresponding author}

\begin{abstract}
The capabilities of recurrent neural networks and Koopman-based frameworks are assessed in the prediction of temporal dynamics of the low-order model of near-wall turbulence by Moehlis {\it et al.} (New J. Phys. {\bf 6}, 56, 2004). Our results show that it is possible to obtain excellent reproductions of the long-term statistics and the dynamic behavior of the chaotic system with properly trained long-short-term memory (LSTM) networks, leading to relative errors in the mean and the fluctuations below $1\%$. Besides, a newly developed Koopman-based framework, called Koopman with nonlinear forcing (KNF), leads to the same level of accuracy in the statistics at a significantly lower computational expense. Furthermore, the KNF framework outperforms the LSTM network when it comes to short-term predictions. We also observe that using a loss function based only on the instantaneous predictions of the chaotic system can lead to suboptimal reproductions in terms of long-term statistics. Thus, we propose a model-selection criterion based on the computed statistics which allows to achieve excellent statistical reconstruction even on small datasets, with minimal loss of accuracy in the instantaneous predictions.
\end{abstract}

\begin{keyword}
Dynamical systems \sep Machine learning \sep Data-driven modeling \sep Recurrent neural networks \sep Koopman operator
\end{keyword}

\end{frontmatter}


\section{Introduction}
\label{sec-introduction}
The potential of machine-learning methods in a wide range of areas~\citep{jean_et_al_2016,defauw_et_al_2018,norouzzadeh_et_al_2018,ham_et_al_2019,udrescu,vinuesa_et_al_2020} has motivated its recent use in the context of fluid mechanics, as discussed for instance by \cite{jimenez_ml}, \cite{duraisamy_et_al} and \cite{brunton_et_al_2020}. Neural networks (NNs), which are computational frameworks used to learn certain tasks from examples, are a widely used tool in machine learning. Their success in a number of applications, mainly related to pattern recognition, can be attributed to the increase in available computational power (through graphics processing units, {\it i.e.} GPUs) and training data which explains the increasing interest in their use for turbulence~\citep{kutz}. Several studies have explored the possibility of using neural networks to develop more accurate Reynolds-averaged Navier--Stokes (RANS) models \citep{ling_et_al,wu_et_al}, while other studies  aim at developing subgrid-scale (SGS) models for large-eddy simulations (LESs) of turbulent flows \citep{lapeyre_et_al,beck_et_al}. On the other hand, NNs have been used for non-intrusive sensing of turbulent flows~\citep{guemes,guastoni,guastoni2}, for the development of efficient flow-control strategies~\citep{rabault}, and to model the near-wall region of wall-bounded turbulence~\citep{milano_koumoutsakos}. Other relevant applications of neural networks include the development of robust inflow conditions for high-Reynolds-number turbulence simulations \citep{fukami_et_al}, super-resolution reconstruction~\citep{fukami2019} and pattern identification in flow data~\citep{raissi_et_al}.

On the other hand, data-driven finite-dimensional approximations of the Koopman operator have also received attention in recent years, in particular, for problems dealing with complex spatiotemporal behavior such as turbulent flows~\citep{Arbabi,giannakis2018,page2019}. Koopman operator theory is an alternative operator-based perspective to dynamical systems theory, which provides a versatile framework for the data-driven study of nonlinear systems. The theory is grounded on the work by \cite{Koopman} and \cite{Koopman2}, and its potential in data-driven analysis of dynamical systems was assessed in the works by \cite{Mezic2004} and \cite{Mezic2005}. The so-called \textit{Koopman operator} is an infinite-dimensional linear operator acting on Hilbert space of observable functions of the state of the system, which describes the evolution of a dynamical system in time. The spectral decomposition of the Koopman operator developed by \cite{Mezic2005} provides useful insight into the underlying dynamics of the nonlinear system and allows to employ traditional techniques in numerical linear algebra for nonlinear systems. In particular, Koopman modes offer a set of coherent structures useful for studying the evolution of the system and to identify the dominant patterns in the data. Modal decomposition of the Koopman operator has been utilized for analysis of complex systems in various engineering fields, including fluid dynamics \citep{rowley}, neuroscience \citep{neuroscience}, robotic control \citep{robotics}, image processing \citep{MultiresolutionDMD}, and system identification \citep{Mauroy}.

The aims of the present work are to assess the potential of NNs and Koopman frameworks to predict the temporal dynamics of a low-order model of turbulent shear flows, and to test various strategies to improve such predictions. In order to easily obtain sufficient data for training and validation, we considered a low-order representation of near-wall turbulence, described by the model proposed by \cite{moehlis_et_al}. The mean profile, streamwise vortices, the streaks and their instabilities as well as their coupling are represented by nine spatial modes $\mathbf{u}_{j}(\mathbf{x})$. The spatial coordinates are denoted by $\mathbf{x}$ and $t$ represents time. The instantaneous velocity fields can be obtained by superimposing the nine modes as: $\mathbf{u}_{{\rm inst}}(\mathbf{x},t) = \sum_{j=1}^9 a_{j}(t) \mathbf{u}_j(\mathbf{x})$, where Galerkin projection can be used to obtain a system of nine ordinary differential equations (ODEs) for the nine mode amplitudes $a_{j}(t)$. Each of the ODEs involves a linear term and several nonlinear terms, all of the form of quadratic nonlinearities ($q_{k}(t) = a_{i}(t)a_{j}(t)$), and they may include a constant, which can be written as:
\begin{equation}
    \dfrac{{\rm d} \mathbf{a}(t)}{{\rm d}t} = \mat{L}\mathbf{a}(t) + \mat{N}\mathbf{q}(t) + \mathbf{c},
    \label{Eq:dyna_sys}
\end{equation}
where $\mathbf{a} \in \mathbb{R}^{n}$, is the vector of mode amplitudes, $\mathbf{q} \in \mathbb{R}^{m}$ is the vector of nonlinear processes, $\mat{L} \in \mathbb{R}^{n \times n}$ and $\mat{N} \in \mathbb{R}^{m \times n}$ are coefficient matrices of linear and nonlinear terms, respectively, and $\mathbf{c} \in \mathbb{R}^{n}$ is the vector of constant values. A model Reyonlds number $Re$ can be defined in terms of the channel full height $2h$ and the laminar velocity $U_{0}$ at a distance of $h/2$ from the top wall. Here we consider $Re=400$ and employ $U_{0}$ and $h$ as velocity and length scales, respectively. The ODE model was used to produce over 10,000 time series of the nine amplitudes, each with a time span of 4,000 time units, for training and validation. The domain size is $L_{x}=4 \pi$, $L_{y}=2$ and $L_{z}=2 \pi$, where $x$, $y$ and $z$ are the streamwise, wall-normal and spanwise coordinates respectively, and we consider only time series that are chaotic over the whole time span. In the next sections we will discuss the feasibility of using various data-driven approaches to predict the temporal dynamics of this simplified turbulent flow. Note that we do not construct a reduced-order model (ROM), since we do not perform model reduction. All the neural-network-based results discussed in this study were obtained using the machine-learning software framework developed by Google Research called TensorFlow~\citep{tensor_flow}. The results from the Koopman-based frameworks were obtained through an in-house implementation of the methods.

This article is organized as follows: in $\S$\ref{sec_NNs} we provide an overview of the predictive capabilities of recurrent neural networks and we summarize some of our previous results; in $\S$\ref{sec_Koopman} we discuss the theoretical background relevant to the Koopman-based frameworks under consideration in this work; the predictive capabilities of both data-driven approaches are compared in $\S$\ref{sec_comparison}; the robustness of the Koopman-based framework with nonlinear forcing is investigated in $\S$\ref{sec_KNF_robustness}; possible ways of improving the performance of recurrent neural networks are discussed in $\S$\ref{sec_NNs_improve}; and finally, in $\S$\ref{sec_conclusions} we provide a summary and the conclusions of the study.

\section{Predictions with recurrent neural networks}
\label{sec_NNs}
The simplest type of neural network is the so-called multilayer perceptron (MLP)~\citep{rumelhart1985learning}, which consists of two or more layers of nodes (also denoted by the term neurons or units), where each node is connected to the ones in the preceding and succeeding layers. Although MLPs are used in practice, their major limitation is that they are designed for point prediction as opposed to time-series prediction, which might require a context-aware method. Nevertheless, MLPs provide a solid baseline in machine-learning applications and thereby help verifying the need for a more sophisticated network architecture. In a previous study~\citep{srinivasan} we assessed the accuracy of MLP predictions of the nine-equation model by \cite{moehlis_et_al}, where the time evolution of the nine coefficients was predicted with several different architectures. The long-term statistics were obtained by averaging over the periodic directions ({\it i.e.}\ $x$ and $z$) and in time over 500 complete time series, which was sufficient to ensure statistical convergence in this case. In order to quantify the accuracy of the predictions, we will consider the relative error between the model and the MLP prediction (denoted by the subindices `mod' and `pred', respectively) for the mean flow as:
\begin{equation}
E_{\overline{u}}=\frac{1}{2\  {\rm max}(\overline{u}_{{\rm mod}})} \int_{-1}^{1} \left | \overline{u}_{{\rm mod}}-\overline{u}_{{\rm pred}} \right |  {\rm d}y,
\end{equation}
where the normalization with the maximum of $\overline{u}$ is introduced to avoid spurious error estimates close to the centerline where the velocity is 0. This error is defined analogously for the streamwise velocity fluctuations $\overline{u^{\prime 2}}$, where the Reynolds decomposition is defined as $u = \overline{u} + u^{\prime}$. Note that the same approach will be used in this study to compute statistics and assess the accuracy of the statistics reproductions. A number of MLP architectures were investigated \citep[see additional details in the work by][]{srinivasan}, and the best predictions were obtained when considering $l=5$, $n=90$ and $p=500$, which denote respectively the number of hidden layers, the number of neurons per layer and the number of previous $a_{j}(t)$ values used to obtain a prediction. With this architecture, the errors in the mean and fluctuations are $E_{\overline{u}}=3.21\%$ and $E_{\overline{u^{\prime 2}}}=18.61\%$ respectively, indicating that although acceptable reproductions of the mean flow can be obtained, the errors in the fluctuations are high. Furthermore, the size of the input was $d=9p=4,500$ ({\it i.e.} 9 coefficients over the past 500 time steps are used to predict the next 9 coefficients), which is quite large. Since the MLP performs point predictions, it does not exploit the sequential nature of the data, and it is therefore important to assess the feasibility of using other types of networks, {\it i.e.} the so-called recurrent neural networks (RNNs), which can benefit from the information contained by the temporal dynamics in the data.

\begin{figure*}[h]
    \centering
	\includegraphics[width=4.5in]{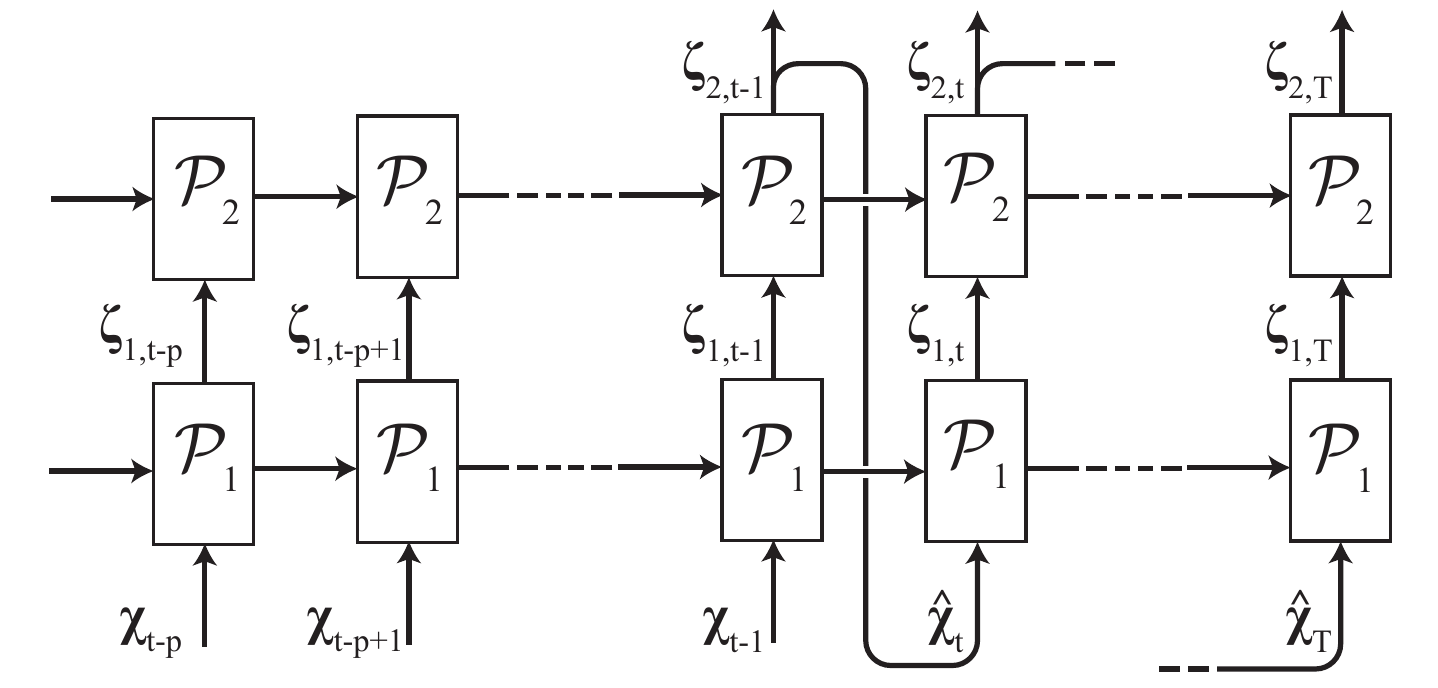}
	\caption{``Unrolled" representation of a multi-layer LSTM, where $\mathcal{P}_i$ is the set of parameters that characterize the LSTM unit of the \textit{i}-th layer. Note that $\mathcal{P}_i$ is shared among all the $p$ time steps considered for the prediction. Here $\pmb{\chi}$ is the network input, $\pmb{\hat{\chi}}$ is the output predicted by the neural network and $T$ is the final time step of the prediction. Note that during training we consider $T=1$ (\textit{i.e.} we predict the coefficients only at one timestep), whereas during testing $T=4000$, and the predictions are made using also the previous predictions from the neural network. This figure does not represent the complete network considered in this work, since the output of the LSTM $\pmb{\hat{\chi}}$ needs to go through an additional fully-connected layer in order to obtain the 9 model coefficients.}
	\label{figure2}
\end{figure*}

In its simplest form, an RNN is a neural network containing a single hidden layer with a feedback loop. As opposed to MLPs, each node of the RNN layer has an internal state vector that is combined with the input vector to compute the output. The output of the hidden layer in the previous time instance is fed back into the hidden layer along with the current input. This allows information to persist, making the network capable of learning sequential dependencies. In practice, this simple recurrent network is not effective to learn long-term dependencies, hence a more sophisticated model is required, such as the long-short-term memory (LSTM) network proposed by \cite{hochreiter_schmidhuber}, or the gated recurrent unit (GRU) network developed by \cite{cho}. Both architectures use a gating mechanism to actively control the dynamics of the recurrent connections. Each unit in the LSTM layer performs four operations through three different gates. The \textit{forget gate} uses the output in the previous time instance $\pmb{\zeta}_{t-1}$ and the current input $\pmb{\chi}_{t}$ to determine which part of the cell state $\mathbf{C}_{t-1}$ should be retained in the current evaluation. The \textit{input gate} uses the same quantities to determine which values of the cell state should be updated and it also computes the candidate values for the update. Finally the \textit{output gate} uses the newly-updated cell state to compute the output. Algorithm~\ref{algo_rnn} illustrates how the output is computed and how the cell state is updated, where $\otimes$ indicates the Hadamard product and $\sigma$ denotes the logistic sigmoid function. A schematic representation of a multi-layer LSTM is shown in \Cref{figure2}, when multiple LSTM layers are stacked, the output of the $i^{th}$ layer $\pmb{\zeta}_{i,t}$ is fed as input to the following LSTM layer. The model is defined by a set of parameters $\mathcal{P}$ which comprise the weight matrices $\mathbf{W}$ and the biases $\mathbf{b}$. During training, the values of the parameters are optimized to minimize a certain loss function.
\begin{algorithm}[h]
\DontPrintSemicolon
\KwIn{Sequence $\pmb{\chi}_{1}, \pmb{\chi}_{2}, \dots \pmb{\chi}_{p}$}
\KwOut{Sequence $\pmb{\zeta}_{1}, \pmb{\zeta}_{2}, \dots \pmb{\zeta}_{p}$}
 set $\mathbf{h}_0 \leftarrow 0$\;
 set $\mathbf{C}_0 \leftarrow 0$\;
 \For{$t\leftarrow 1$ \KwTo $p$}{
  \ \ $\mathbf{f}_t \leftarrow \sigma(\mathbf{W}_{f} [\pmb{\chi}_{t}, \pmb{\zeta}_{t-1}] + \mathbf{b}_{f})$\;
  \ \ $\mathbf{i}_t \leftarrow \sigma(\mathbf{W}_{i} [\pmb{\chi}_{t}, \pmb{\zeta}_{t-1}] + \mathbf{b}_{i})$\;
  \ \ $\mathbf{\widetilde{C}}_t \leftarrow \tanh(\mathbf{W}_{f} [\pmb{\chi}_{t}, \pmb{\zeta}_{t-1}] + \mathbf{b}_{f})$\;
  \ \ $\mathbf{C}_t \leftarrow \mathbf{f}_t\otimes\mathbf{C}_{t-1} + \mathbf{i}_t\otimes\mathbf{\widetilde{C}}_t$\;
  \ \ $\mathbf{o}_t \leftarrow \sigma(\mathbf{W}_{o} [\pmb{\chi}_{t}, \pmb{\zeta}_{t-1}] + \mathbf{b}_{o})$\;
  \ \ $\pmb{\zeta}_t \leftarrow \mathbf{o}_t\otimes\tanh(\mathbf{C}_{t-1})$\;
 }
 \caption{Compute the output sequence of an LSTM network.}
 \label{algo_rnn}
\end{algorithm}

In our previous work~\citep{srinivasan} we analyzed the prediction capabilities of LSTM networks for this low-order turbulent shear flow wall model by considering a network with a single layer of 90 LSTM units, where the number of units commonly refers to the dimensionality of the output space of the LSTM, similarly to what is done with MLPs. Since the LSTM layer output is $\pmb{\zeta}_t \in \mathbb{R}^{90}$, a fully-connected layer is added to obtain the prediction of the 9 model coefficients $a_j(t)$. We trained it with three different datasets, consisting respectively of 100, 1,000, and 10,000 time series spanning 4,000 time units each. We considered a validation loss defined as the  sum  over $p$ time steps of the squared error in the prediction of the instantaneous coefficients $a_{j}$, and observed that better predictions could be obtained when larger datasets were employed for training. Note that we considered $20\%$ of the training data as a validation set, which is used to check the evolution of the loss on data which has not been seen by the network during training. Using 10,000 time series for training, we obtained accurate predictions of the long-term statistics, with $E_{\overline{u}}=0.45\%$ and $E_{\overline{u^{\prime 2}}}=2.49\%$. This was obtained with $p=10$, {\it i.e.} with an input size 50 times smaller than that used with the MLP. The agreement of all the statistics with the reference data was compelling, and even higher-order moments exhibited low relative errors, {\it i.e.}  $1.01\%$ and $2.57\%$ for skewness and flatness, respectively~\citep{srinivasan}. These results highlight the excellent predicting capabilities of the LSTM network, given that sufficient training data is provided, due to the ability of the network to exploit the sequential nature of the data.

\section{Koopman-based frameworks}
\label{sec_Koopman}
Predicting the spatiotemporal evolution of high-dimensional and nonlinear dynamical systems (such as turbulent flows) based on finite-dimensional approximations of the Koopman operator is of particular interest for fluid mechanics. Dynamic Mode Decomposition (DMD) \citep{Schmid2010,Tu2014} is one of the most popular algorithms for modal decomposition based on the Koopman operator \citep{rowley}. DMD, in its original formulation, implicitly utilizes linear observables of the state of the system. However, linear functions may not be rich enough to describe many nonlinear dynamical systems. The Extended DMD (EDMD) \citep{Williams2015} was proposed to include a richer set of nonlinear observable functions (such a set is denoted as {\it dictionary}) for better approximations of the Koopman eigenfunctions. Through a careful choice of the dictionary, it was shown that the EDMD algorithm has better performance than DMD \citep{Williams2015}. However, a drawback of the EDMD algorithm is the fact that, without a priori knowledge about the underlying dynamics, it is not clear how to choose a dictionary that is sufficiently rich to span a useful Koopman-invariant subspace. Recently, several studies have introduced fully data-driven approaches for learning Koopman embedding and autonomous dictionary learning using Deep Neural Networks (DNNs): \cite{Takeishi2017,Li2017,Lusch2018}.

The necessity of choosing appropriate input data is critical for data-driven modeling and prediction of dynamical systems using Koopman-based frameworks. Instead of utilizing linear or nonlinear observable functions of the state variables, it may be possible to construct a rich feature space using delay-embedding of time series measurements. Time delay-embedding, also known as delay-coordinate embedding, is based on Takens embedding theorem \citep{Takens} and refers to the inclusion of previous data in dynamical system models. Time-delay embedding has been widely used for state space reconstruction and analysis of chaotic systems \citep{Farmer1987,Crutchfield1987,Abarbanel1993,Sugihara496}. By combining delay embedding with DMD, \cite{Arbabi2017} introduced the Hankel-DMD method, which is a linear model that can provide a representation of the Koopman eigenvalues and eigenfunctions. 

\cite{Brunton2017} presented a universal data-driven decomposition of chaos as an intermittently forced linear system. Their model, referred to as Hankel alternative view of Koopman (HAVOK), combines Takens’ delay embedding with modern Koopman-operator theory and sparse regression to obtain linear representations of strongly nonlinear dynamics. \cite{Brunton2017} applied this model to the canonical Lorenz system, as well as to real-world examples of chaos leading to accurate prediction of attractor switching and bursting phenomena in such cases. More recently, \cite{khodkar2019} introduced a successful Koopman-based framework for data-driven spatiotemporal prediction of high-dimensional and highly chaotic systems. The main novelty of their approach is the fact that the nonlinearities are modeled through external forcing, where the observables are vector-valued and delay-embedded. The model has been shown capable of accurate prediction of well-known prototypes of chaos, such as the Kuramoto-Sivashinsky equation~\citep{Kuramoto,Sivashinsky} and the Lorenz-96 system~\citep{lorenz96}, as well as high-Reynolds-number lid-driven cavity flows \citep{Arbabi} for several Lyapunov timescales.

In this work, we leverage the recent advances in the prediction of chaotic systems using Koopman-based frameworks for the prediction of a low-order model of turbulent shear flows. In particular, we utilize the method introduced by \cite{khodkar2019} for time series prediction of the nine-equation model. 

\subsection{Koopman-operator theory}

The Koopman operator theory is central to all that follows in this section, therefore we provide a brief overview of the mathematical aspects and definition of the properties relevant to our study. To this end, we focus on an autonomous discrete-time dynamical system:
\begin{equation}
    \mathbf{x}_{t+1} = \mathbf{F}(\mathbf{x}_{t}),
\end{equation}
on the state space $\mathcal{M}\subseteq \mathbb{R}^{n}$ , where $\mathbf{x}$ is a coordinate vector of the state, and $\mathbf{F}:\mathcal{M} \to \mathcal{M}$ is the evolution operator. The Koopman operator $\mathbfcal{K}$ is defined as an infinite-dimensional linear operator that acts on functions of state space (\textit{observables}) $g:\mathcal{M}\to\mathbb{C}$ (unlike $\mathbf{F}$, which acts on $\mathbf{x}\in\mathcal{M}$). The action of the Koopman operator is:
\begin{equation}
    \mathbfcal{K} g = g \circ\mathbf{F},
\end{equation}
where $\circ$ indicates the composition of $g$ with $\mathbf{F}$. 
In fact, the Koopman operator defines a new infinite-dimensional linear dynamical system that governs the evolution of the \textit{observables} $g_{t} = g(\mathbf{x}_t)$ in discrete time. Note that $\mathbfcal{K}$ is infinite-dimensional even if $\mathbf{F}$ is finite-dimensional, and also it is linear even when $\mathbf{F}$ is nonlinear. For a detailed discussion of the Koopman operator, the readers are referred to the available research articles \citep{Mezic2005,rowley} and reviews on the topic \citep{AppliedKoopmanism,Mezic2013}.

\subsection{Koopman-based framework with nonlinearities modeled as exogenous \\ forcing}

Obtaining finite-dimensional approximations of the Koopman operator is the focus of intense research efforts due to its capabilities when it comes to linear representation of the nonlinear dynamical systems. This is also related to the wealth of methods available for estimation, prediction, and control of linear systems. The Hankel DMD algorithm \citep{Arbabi2017} provides a practical numerical framework for computation of the Koopman spectrum by applying DMD to the so-called Hankel matrix of data $\mathbfcal{H}$: 
\begin{equation}
    \label{H_matrix}
    \mathbfcal{H} = 
    \begin{bmatrix}
        \mathbf{x}^{1} &  \mathbf{x}^{2} & \cdots &  \mathbf{x}^{N-q+1}\\
        \mathbf{x}^{2} &  \mathbf{x}^{3} & \cdots &  \mathbf{x}^{N-q+2} \\
        \vdots & \vdots & \ddots & \vdots \\
        \mathbf{x}^{q} &  \mathbf{x}^{q+1} & \cdots &  \mathbf{x}^{N}
    \end{bmatrix}, \\
 \end{equation}
 where $N$ is the number of the vector-valued observables $\mathbf{x}^{i}$ sampled at $t = i \tau$, $\tau$ is the sampling interval and $q$ is the delay-embedding dimension. For the nine-equation model of \cite{moehlis_et_al}, $\mathbf{x}^{i}$ is equal to $\begin{bmatrix}
    a_{1}^{i}& a_{2}^{i}& \cdots & a_{9}^{i}     
 \end{bmatrix}^{\mathsf{T}}$ in \cref{H_matrix}, where $\mathsf{T}$ indicates transpose.  Therefore, $\mathbfcal{H}$ is a matrix with the size of $(n \times q) \times (N-q+1)$, where $n$ is the number of state variables. Following the Exact DMD algorithm formulation~\citep{Tu2014,Arbabi2017}, we define:
 \begin{equation}
    \mat{X} = \begin{bmatrix}
        \mathbfcal{X}^{1} & \cdots & \mathbfcal{X}^{N-q}
    \end{bmatrix}
    ,~\mat{Y} = \begin{bmatrix}
        \mathbfcal{X}^{2} & \cdots & \mathbfcal{X}^{N-q+1}
    \end{bmatrix},
 \end{equation}
 where $\mathbfcal{X}^{i}$ denotes the $i^{th}$ column of the Hankel matrix $\mathbfcal{H}$. The Singular Value Decomposition (SVD) of matrix $\mat{X}$ is computed as:
 \begin{equation}
    \mat{X} = \mat{USV}^{*},
\end{equation}
where $^{*}$ denotes the conjugate transpose, $ \mat{U} \in \mathbb{C}^{(n \times q) \times r} $, $ \mat{S} \in \mathbb{C}^{r \times r}$ , and $ \mat{V} \in \mathbb{C}^{(N-q) \times r}$. Here, $r$ is the rank of the reduced SVD approximation to $\mat{X}$. The finite-dimensional approximation of the Koopman operator using Hankel-DMD (HDMD) is computed as:
\begin{equation}
    \mat{\tilde{A}}_{\mathrm{HDMD}} = \mat{U}^{*}\mat{YVS}^{-1},
\end{equation}
with size $ r \times r$. Once the HDMD operator $\mat{\tilde{A}}_{\mathrm{HDMD}}$ is calculated using the training set, a future vector-valued observable $\mathbf{x}^{m+1}$ can be predicted from:
\begin{equation}
    \mathbf{X}^{m+1, r} = \mat{\tilde{A}}_{\mathrm{HDMD}} \mathbf{X}^{m, r},
\end{equation}
where $\mathbf{X}^{i, r}$ $($of size $r \times 1)$ is the projection of $\mathbf{X}^{i} = \begin{bmatrix}
    \mathbf{x}^{i-q}&\mathbf{x}^{i-q+1}&\cdots&\mathbf{x}^{i}
\end{bmatrix}^{\mathsf{T}}$, which has size $(n \times q) \times 1$, onto the subspace of first $r$ singular vectors.

\cite{khodkar2019} showed that the linear combination of a finite number of DMD modes may not be sufficient to obtain an accurate representation of the long-term nonlinear characteristics of a chaotic dynamical system such as the Kuramoto-Sivashinsky equation~\citep{Kuramoto,Sivashinsky}. On the other hand, the HAVOK model introduced by \cite{Brunton2017} has been shown to provide excellent reproductions of nonlinear dynamics by adding a forcing term to the linear model. Inspired by the HAVOK model, \cite{khodkar2019} proposed a new Koopman-based framework, which incorporates nonlinear effects through external forcing, so a dynamical system can be modeled as:
\begin{equation}
    \label{Eq_m2}
    \mathbf{x}^{m+1} = \mat{A} \mathbf{x}^m + \mat{B} \mathbf{f}^m,
\end{equation}
where $\mathbf{f}$ denotes the forcing term. We construct $\mathbf{f}$ as an augmented vector library, containing any candidate nonlinear functions of $\mathbf{x}$, \emph{e.g.}, polynomial terms:
\begin{equation}
    \mathbf{f}^{i} = \begin{bmatrix}
        1& (\mathbf{x}^{i})^{p_{2}}& (\mathbf{x}^{i})^{p_{3}} & \cdots & (\mathbf{x}^{i})^{p_{n}}
    \end{bmatrix}^\mathsf{T},
\end{equation}
where, for instance, $(\mathbf{x}^{i})^{p_{2}}$ and $(\mathbf{x}^{i})^{p_{3}}$ indicate any possible quadratic and cubic nonlinearities, respectively ($a^{i}_{j}a^{i}_{k}$ and $a^{i}_{j}a^{i}_{k}a^{i}_{l}$ where $j,k,l \in [1, 9]$). A constant can also be considered in this vector. For many systems of interest, the vector $\mathbf{f}$ consists of only few terms, making it sparse in the space of possible functions. To identify forms of nonlinearities in the data we implement an iterative procedure from the sparse identification of nonlinear dynamics (SINDy) method proposed by \cite{Brunton2016}. In this procedure, which is presented in \Cref{algo:ridge}, we start with an iterative linear regression of $\begin{bmatrix}
    \mathbf{x}^{2} & \mathbf{x}^{3} & \cdots & \mathbf{x}^{N}
\end{bmatrix}$ on $\begin{bmatrix}
    \mathbf{xf}^{1} & \mathbf{xf}^{2} & \cdots & \mathbf{xf}^{N-1}
\end{bmatrix}$ where $ \mathbf{xf}^{i} = \begin{bmatrix}
    \mathbf{x}^{i} & \mathbf{f}^{i}
\end{bmatrix}^{\mathsf{T}}$  and then zero out all coefficients that are smaller than a threshold $\varepsilon$. Then, we perform another iterative linear regression using only the terms that have non-zero coefficients. These new coefficients are again zeroed using the threshold, and the procedure is continued until the non-zero coefficients converge. At the end, the nonlinear terms with non-zero coefficients are the identified nonlinearities from the original data. We used the ridge regression, which is a regularized linear-regression technique, to find the coefficients. It leads to the learned coefficients that are biased toward zero due to the $\ell_{2}$ regularization and makes the process of the elimination of small coefficients more efficient in practice.

\SetKwComment{Comment}{$\triangleright$\ }{}
\begin{algorithm}[h]
    \SetKw{KwInit}{Initialize}
    \DontPrintSemicolon
    \KwIn{$\mat{y} = \begin{bmatrix}
        \mathbf{x}^{2} & \mathbf{x}^{3} & \cdots & \mathbf{x}^{N}
    \end{bmatrix},~\mat{x} = \begin{bmatrix}
        \mathbf{xf}^{1} & \mathbf{xf}^{2} & \cdots & \mathbf{xf}^{N-1}
    \end{bmatrix}$, threshold value $\varepsilon$}
    \KwOut{$\mathbf{I}_{\mathrm{active}}$ \Comment*[r]{Indices of active nonlinearities}}
    \vspace{0.05in}
    $n, m \leftarrow$ number of rows of $\mat{y}, \mat{x}$\;
    \KwInit{$\mathbf{C}(n, m)$ \Comment*[r]{Coefficients}}
    \KwInit{$\mathbf{I}(n, m)$ \Comment*[r]{Active indices, dtype = bool}}
    \vspace{0.05in}
     \For{$i\leftarrow 1$ \KwTo $\mathrm{Max~Iteration = 20}$}{
         \For{$j\leftarrow 1$ \KwTo $n$}{
             $\mathbf{I}_j \leftarrow \mathbf{I}[j, ~:]$\;
            $\mathbf{C}[j,~\mathbf{I}_j] \leftarrow \mathrm{Ridge~Regression}(\mat{y}[j,~:], \mat{x}[\mathbf{I}_j ,~:])$\;
            $\mathbf{I}_j \leftarrow \mathrm{abs}(\mathbf{C}[j,~:]) >= \varepsilon$ \Comment*[r]{Find big coefficients}
            $\mathbf{C}[j,~\sim\mathbf{I}_j] \leftarrow 0$ \Comment*[r]{Zero out small coefficients}
            $\mathbf{I}[j,~:]  \leftarrow \mathbf{I}_j$
         }
        \If{$\mathbf{C}$ \rm{does not change}}{
            \textbf{Break}\;
        }
     }
    $\mathbf{I}_{\mathrm{active}} \leftarrow$ \rm{Maximum element of each column in }$\mathbf{I}[:,~n:]$\;
     \caption{Compute sparse candidate nonlinear terms for the KNF method.}
     \label{algo:ridge}
\end{algorithm}

Once the forms of nonlinearities in the data are identified, we construct $\mathbf{f}$ using these terms and mask out the others.
We define the time-delay-embedded form of \cref{Eq_m2} as: 
\begin{equation}
    \label{Eq_m2_embedded}
    \mathbf{X}^{m+1} = \mat{A} \mathbf{X}^m + \mat{B} \mathbfcal{F}^m,
\end{equation}
where $\mathbf{X}^{m}$ is the same as above, and $\mathbfcal{F}$ is the Hankel representation of the forcing vectors:
\begin{equation}
    \label{F_matrix}
    \mathbfcal{F} = 
    \begin{bmatrix}
        \mathbf{f}^{1} &  \mathbf{f}^{2} & \cdots &  \mathbf{f}^{N-q}\\
        \mathbf{f}^{2} &  \mathbf{f}^{3} & \cdots &  \mathbf{f}^{N-q+1} \\
        \vdots & \vdots & \ddots & \vdots \\
        \mathbf{f}^{q} &  \mathbf{f}^{q+1} & \cdots &  \mathbf{f}^{N-1}
    \end{bmatrix}. \\
 \end{equation}
 The size of $\mathbfcal{F}$ is $(n^{\prime} \times q) \times (N-q)$, where $n^{\prime}$ is the size of the forcing vector $\mathbf{f}$ and it depends on the form of the nonlinearities and the number of nonlinear processes in the dynamical system. 

 The unknown maps of $\mat{A}$ and $\mat{B}$ can be found using the DMDc algorithm (where $c$ stands for control) introduced by \cite{Proctor2016} as:
 \begin{equation}
    \centering
    \begin{array}{l}
        \mat{A} = \mat{\hat{U}}^{*}\mat{Y}\mat{\tilde{V}\tilde{S}}^{-1}\mat{\tilde{U}}_{1}^{*}\mat{\hat{U}},\\
        \mat{B} =  \mat{\hat{U}}^{*}\mat{Y}\mat{\tilde{V}\tilde{S}}^{-1}\mat{\tilde{U}}_{2}^{*},
    \end{array}
 \end{equation}
 where $\mat{Y} = \mat{\hat{U}\hat{S}\hat{V}}^{*}$, the truncation rank is $r$ and $\mat{\hat{U}} \in \mathbb{R}^{(n \times q) \times r}$, $\mat{\hat{S}} \in \mathbb{R}^{r \times r}$, and $\mat{\hat{V}} \in \mathbb{R}^{(N - q) \times r}$. Also, $\begin{bmatrix}
     \mat{X} & \mathbfcal{F}
 \end{bmatrix}^{\mathsf{T}} = \mat{\tilde{U}\tilde{S}\tilde{V}}^{*}$, where the truncation rank is $k$ and $\mat{\tilde{U}} \in \mathbb{R}^{((n + n^{\prime}) \times q) \times k}$, $\mat{\tilde{S}} \in \mathbb{R}^{k \times k}$, and $\mat{\tilde{V}} \in \mathbb{R}^{(N - q) \times k}$. Moreover, $\mat{\tilde{U}}_{1} \in \mathbb{R}^{(n \times q) \times k}$ and $\mat{\tilde{U}}_{2} \in \mathbb{R}^{(n^{\prime} \times q) \times k}$ where $\mat{\tilde{U}} = \begin{bmatrix}
    \mat{\tilde{U}}_{1}^{*} & \mat{\tilde{U}}_{2}^{*}
\end{bmatrix}^{\mathsf{T}}$. Here, $\hat{(\cdot)}$ and $\tilde{(\cdot)}$ denote the rank-truncated forms of the SVD matrices from $\mat{Y}$ and $\begin{bmatrix}
    \mat{X} & \mathbfcal{F}
\end{bmatrix}^{\mathsf{T}}$, respectively. Note that $\mat{A}$ and $\mat{B}$ are represented in a reduced-order subspace and have sizes of $r \times r$ and $r \times (n' \times q)$, respectively. Moreover, $r$ and $k$ can be chosen based on SVD rank-truncation methods such as the optimal hard threshold presented by \citep{Gavish2014}. Hereafter, we refer to the Hankel-DMD method introduced by \cite{Arbabi2017} as HDMD and the Koopman-based framework proposed by \cite{khodkar2019} as KNF, which stands for Koopman with nonlinear forcing.

\section{Comparison between predictions from Koopman-based frameworks and LSTM network}
\label{sec_comparison}

In this section, the performance of two Koopman-based frameworks, {\it i.e.} the KNF and HDMD methods, is compared with that of the LSTM network in the prediction of the temporal behavior of the nine-equation model. As discussed in $\S$\ref{sec-introduction}, all the nonlinearities in the nine-equation model are of the form of quadratic nonlinear terms. The KNF model can represent exactly the same form of nonlinearities by proper construction of the forcing term, thus given an accurate choice of the coefficients, it has the potential to exactly reproduce the nine-equation model. Conversely, the LSTM network models nonlinearities through the use of nonlinear activation functions, and the HDMD model seeks a (high-dimensional) linear representation of the nonlinear dynamics. Therefore, the use of the nine-equation model as the reference gives an advantage to the KNF model over the LSTM network in the representation of nonlinearities that should be taken into account in the following comparison. Note however that such advantage is not present when real wall-bounded turbulence data are considered.

\subsection{Short-term predictions}

Here the short-term prediction capabilities of the KNF and HDMD methods are compared with that of the LSTM network. The testing set is the same for all the methods and contains 500 time series, each of them spanning 4,000 time units. The KNF and HDMD models are trained with one set of time series comprising 10,000 time units generated from the nine-equation model. The training time series are checked to be chaotic over the whole time span. The delay-embedding dimension ($q$) is considered equal to 5. Here, we construct the forcing term of the KNF method by including all the possible quadratic, cubic, and quartic terms. We will discuss the robustness of the KNF models in terms of the form of the forcing term in $\S$\ref{sec_KNF_robustness}. Moreover, an LSTM network consisting of one hidden layer with 90 LSTM units and $p = 10$ is also used for the predictions (see \cite{srinivasan} for additional details). The LSTM network is trained with 10,000 sets of time series, each with a time span of 4,000 time units. 

\begin{figure}[h]
    \centering
    \includegraphics[width=4in]{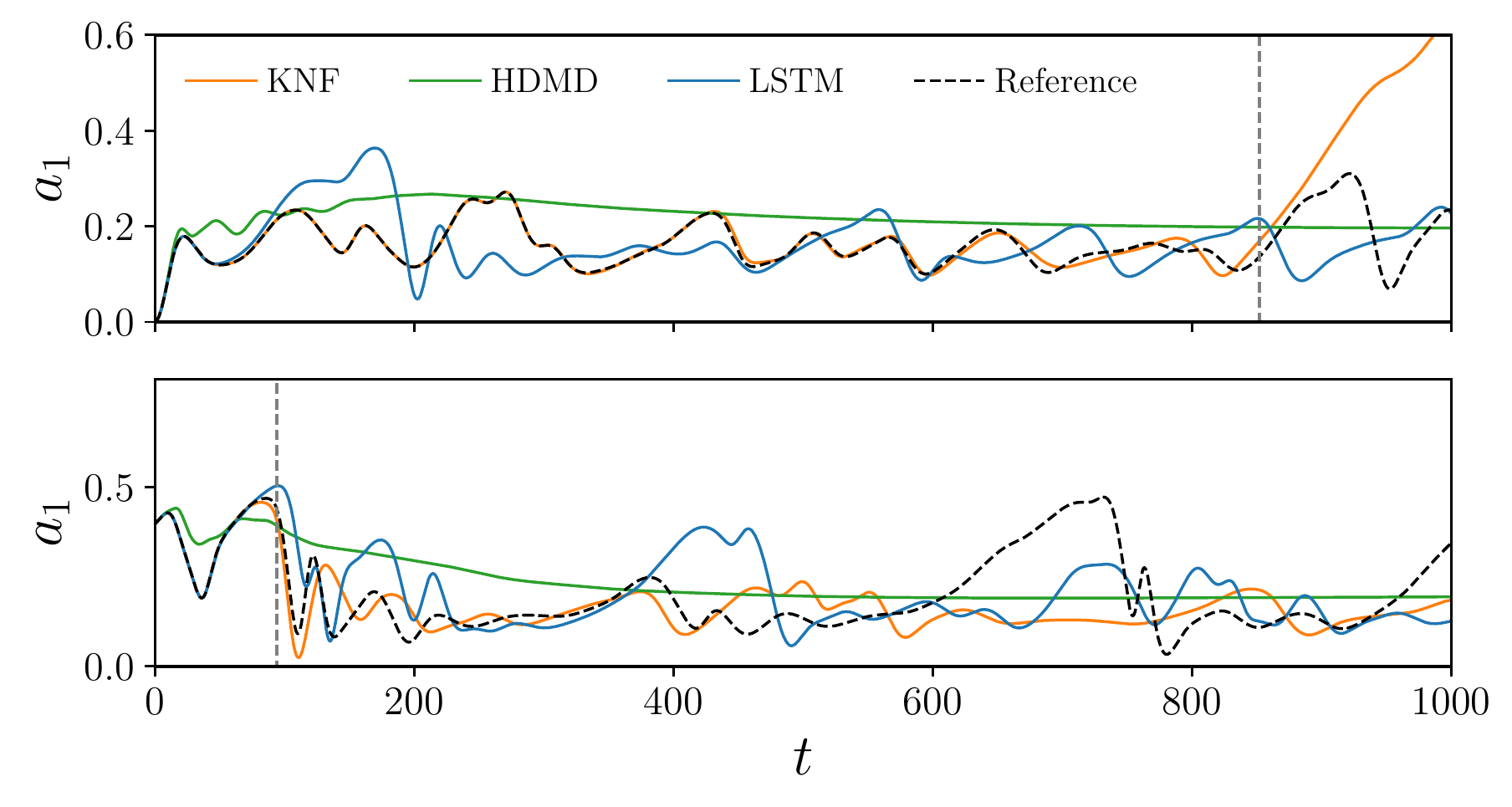}
    \caption{Comparison of short-term prediction capabilities of the first coefficient from the three data-driven approaches. Results are reported for the time series with different initial conditions for which the KNF method provides the longest (top) and the shortest (bottom) prediction horizons. Vertical dashed lines approximately show the prediction horizon of the KNF method, defined as the first point where $\epsilon >0.3 $ for that particular time series.}
    \label{short-term}
\end{figure}

Due to the chaotic nature of the nine-equation model, the performance of the methods depends on the initial condition from which the prediction is conducted. To provide comprehensive insight into the performance of these methods, examples of the predicted trajectories for the amplitude of the first mode $a_{1}(t)$ versus the true trajectory for two specific initial conditions are presented in \Cref{short-term}. Here we show the cases where the KNF method provided the longest (\Cref{short-term}, top) and the shortest (\Cref{short-term}, bottom) intervals in very close agreement with the reference. 
It can be seen in \Cref{short-term}~(top) that the KNF method accurately predicts the time evolution of the $a_{1}$ amplitude for up to $t \simeq 650$, and provides acceptable results for even up to $t \simeq 850$. The LSTM network provides the next best performance, exhibiting accurate predictions of the time evolution for over $t \simeq 90$. On the other hand, \Cref{short-term}~(bottom) shows the case of worst performance from the KNF method, still providing accurate predictions up to $t \simeq 80$ (approximately as long as the LSTM network), and producing acceptable predictions up to $t \simeq 95$. The HDMD method provides accurate predictions for up to $t \simeq 10$ in both cases, and after that, the predictions gradually decay to a constant value. Note that the first 10 and 5 time steps are used to start the predictions for the LSTM network and the Koopman-based models, respectively.
\begin{figure}
    \centering
    \includegraphics[width=4in]{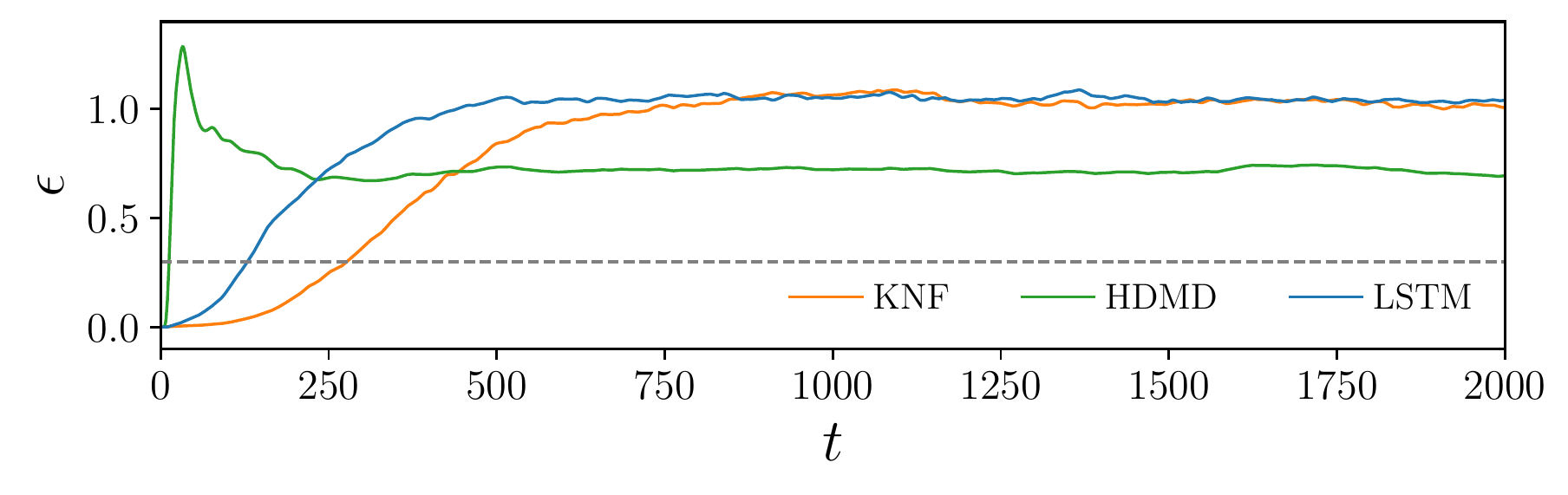}
    \caption{Relative Euclidean norm of errors $\epsilon(t)$ averaged over 500 randomly chosen initial conditions, as defined in \cref{Eq_err_short_term}. The dashed horizontal line shows the threshold value considered for accurate predictions, namely $\epsilon=0.3$. The LSTM network consists of 1 layer and 90 neurons, and it is trained with 10,000 datasets~\citep{srinivasan}. The KNF and HDMD models are trained with one time series, with $N$ = 10,000 and $q$ = 5.}
    \label{Error-short-term}
\end{figure}
We define an averaged relative Euclidean norm of errors $\epsilon(t)$ in nine-dimensional space between the true and predicted trajectories to compare the results over all 500 randomly chosen initial conditions:
\begin{equation}
    \label{Eq_err_short_term}
     \epsilon(t) = \Bigg\langle\dfrac{\left [ \sum \limits_{i=1}^{9} \left (a_{i, {\rm mod}}(t) - a_{i, {\rm pred}}(t)  \right )^{2} \right ]^{1/2}}{ { \big \langle \left[ \sum \limits_{i=1}^{9} \left (a_{i, {\rm mod}}(t) \right )^{2} \right]^{1/2} \big \rangle_{t}}}\Bigg\rangle_{{\rm ens}},
\end{equation}
where $\big \langle \cdot \big \rangle_{{\rm ens}}$ and $\big \langle \cdot \big \rangle_{t}$ indicate ensemble averaging over 500 sets of time series and over 4,000 time units, respectively.~\Cref{Error-short-term} compares $\epsilon(t)$ for the KNF and HDMD methods with the LSTM network. It is evident that the KNF method outperforms the LSTM network for short-term predictions, while it provides the same level of accuracy for long-term predictions. In particular, considering a threshold of $\epsilon=0.3$ to define very good agreement of the coefficient predictions, we observe that the  method provides accurate instantaneous predictions for around 280 time units, while the prediction horizon for the LSTM network is 130 time units. This value is about 12 time units for the HDMD method. Moreover, it can be observed in \Cref{Error-short-term} that for $t > 1000$ the value of $\epsilon(t)$ is around one for the KNF and LSTM, indicating that the variance of the error is as large as the time average of the signal itself, but the predictions have approximately the same mean as the reference amplitudes.

\subsection{Long-term predictions and statistics}
\begin{figure}
    \centering
    \includegraphics[width=4in]{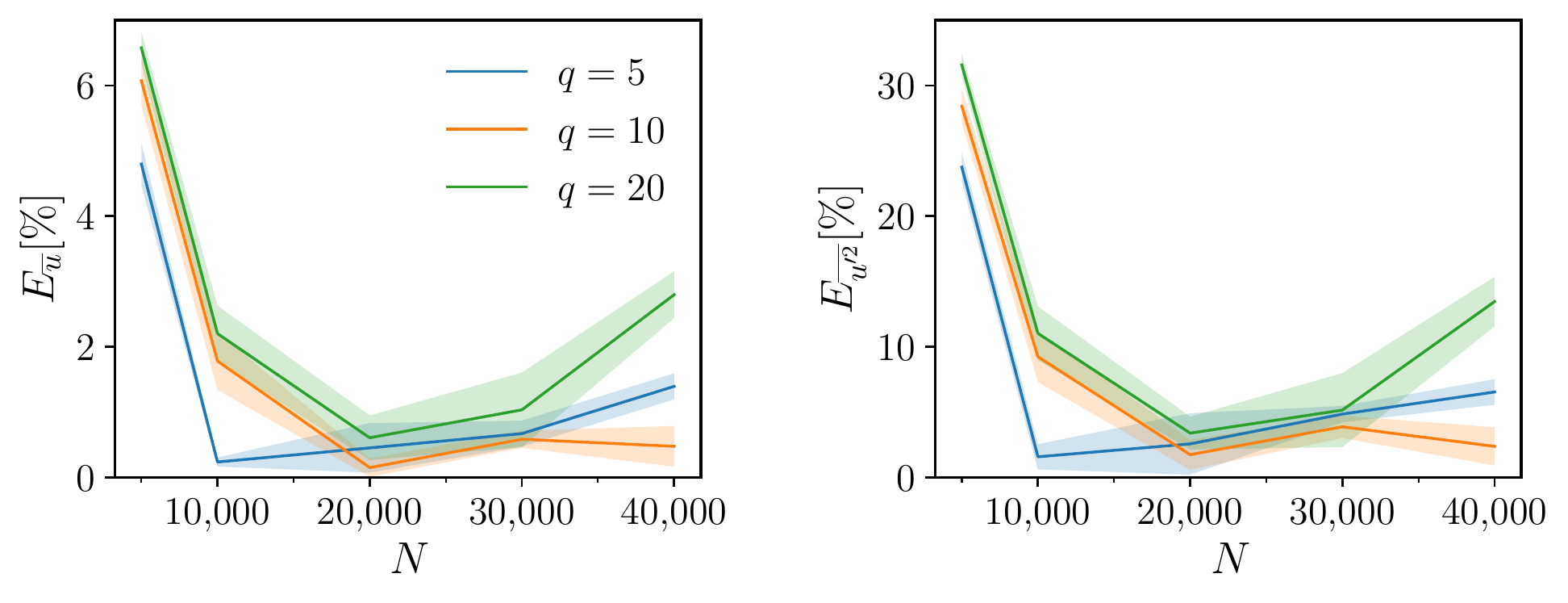}
    \caption{The error in the reproduction of long-term statistics obtained from KNF models using different training data sets with $N$ and $q$ varying from 5,000 to 40,000 and from 5 to 20, respectively. Solid lines represent mean errors over four reference test sets, while the shaded regions correspond to the standard deviation around the mean.}
    \label{fig:knf_table_data}
\end{figure}

We have shown the performance of three data-driven approaches in the short-term prediction of the temporal dynamics of the nine-equation model. Our results indicate that the predictions of all the methods discussed earlier eventually diverge from the true trajectory. However, it is still interesting to examine the performance of the models in the reproduction of the long-term statistical properties of the actual model. Reproduction of the long-term behavior of a chaotic dynamical system using inexpensive data-driven methods can be significantly beneficial in the domain of data-driven turbulence modeling. The predicted mode amplitudes are employed to reconstruct the velocity fields. These fields are used to calculate the long-term statistics by averaging over periodic directions ($x$ and $z$) and over the 4,000 time units of a time series and then performing an ensemble average over 500 different time series of the test set to ensure statistical convergence. Here, we compare the performance of the KNF method and the LSTM network in the reproduction of the long-term dynamics of the nine-equation model. To this end, we first examine the effect of the sizes of the training set $N$ and the delay dimension $q$ on the performance of the KNF method in the reproduction of the long-term statistics, i.e., the mean velocity profile $\overline{u}(y)$ and streamwise velocity fluctuations $\overline{u^{\prime 2}}(y)$. A time series spanning $N$ time units with a delay dimension of $q$ is used to train the KNF method, where $N$ and $q$ vary from 5,000 to 40,000 and from 5 to 20, respectively. The KNF models are trained once, and then the resulting statistics are compared with four different reference test sets (not seen during the training and validation steps). Each of the test sets consists of 500 time series, created from 500 randomly generated initial conditions, which is the difference between the test sets. Each of the time series spans 4,000 time units. We utilized 500 time series to ensure the statistical convergence and four test sets to evaluate the independence of the reproduced statistics to the initial conditions. The results are depicted in \Cref{fig:knf_table_data}.

Our results show that the model trained using a time series spanning 10,000 time units, with a delay dimension of 5, yields the best reproduction of long-term statistics, with mean errors of $E_{\overline{u}} = 0.23\%$ and $E_{\overline{u^{\prime 2}}} = 1.57\%$ over four test sets. It can be seen in \Cref{fig:knf_table_data} that the errors, as expected, are robust for the four different reference test sets. This indicates that it is possible to use a data set as the validation set and find the best set of hyper-parameters, which leads to similar error levels in the testing data set. It can also be observed that utilizing larger training data sets with higher values for delay-embedding dimension may not lead to a more accurate prediction of the long-term statistics. Below we show that the instantaneous predictions improve with an increasing amount of training data, leading to a decrease in the validation loss based on one-step predictions. We observe that using a loss function only based on one-step predictions can lead to overfitting to the instantaneous predictions and suboptimal performance in terms of the reproduction of long-term statistics as can be seen in \Cref{fig:knf_table_data}. We should note that the mapping matrices of $\mat{A}$ and $\mat{B}$ are not sparse; every variable in time-delayed vectors of $\mathbf{X}^{m}$ and $\mathbfcal{F}^{m}$ in \cref{Eq_m2_embedded} takes part in the prediction of each variable at the next time step ($\mathbf{X}^{m+1}$). One possible way to avoid overfitting based on regularization is to use sparsity constraints (\emph{e.g.}, low $\ell_{1}$-norm) on the coefficients of the KNF model. However, in this study, we mainly focused on the role of the size of the training data set in mitigating overfitting issues.

In the next test, we train the KNF model with five different time series to build five different models and reproduce the long-term statistics based on 500 time series to provide error bars of the prediction. These five training time series are generated from five different random initial conditions. $N$ and $q$ are equal to 10,000 and 5, respectively, considering the best performing KNF model in \Cref{fig:knf_table_data}. Results are presented in Table~\ref{KNF_tst}, showing the mean errors of $E_{\overline{u}} = 0.33\%$ and $E_{\overline{u^{\prime 2}}} = 1.40\%$. Very similar error levels are obtained using the other test sets as a reference.

\begin{table}[h]
    \centering
    \caption{Error of long-term statistics for KNF models obtained with 5 different training sets generated from 5 different random initial conditions. In all the cases we consider $N=10,000$ and $q=5$. The best-performing KNF model is highlighted in boldface.}
    \begin{tabular}{ccc}
        \hline \hline
        \rule{0pt}{3ex} Training set & $E_{\overline{u}}[\%]$ & $E_{\overline{u^{\prime 2}}}[\%]$\\
        \vspace{-0.3cm} \\
        \hline
        \vspace{-0.3cm} \\
        1   &   0.44  & 0.77 \\
        \textbf{2}   &   \textbf{0.24}  & \textbf{0.65} \\
        3  &   0.49  &1.54 \\
        4  &   0.32  &2.91 \\
        5  &   0.15  &1.15 \\
        Mean & 0.33 & 1.40 \\
        \vspace{-0.3cm} \\
    \hline \hline
    \end{tabular}
    \label{KNF_tst}
\end{table}

\begin{figure}[h]
    \centering
    \includegraphics[width=4in]{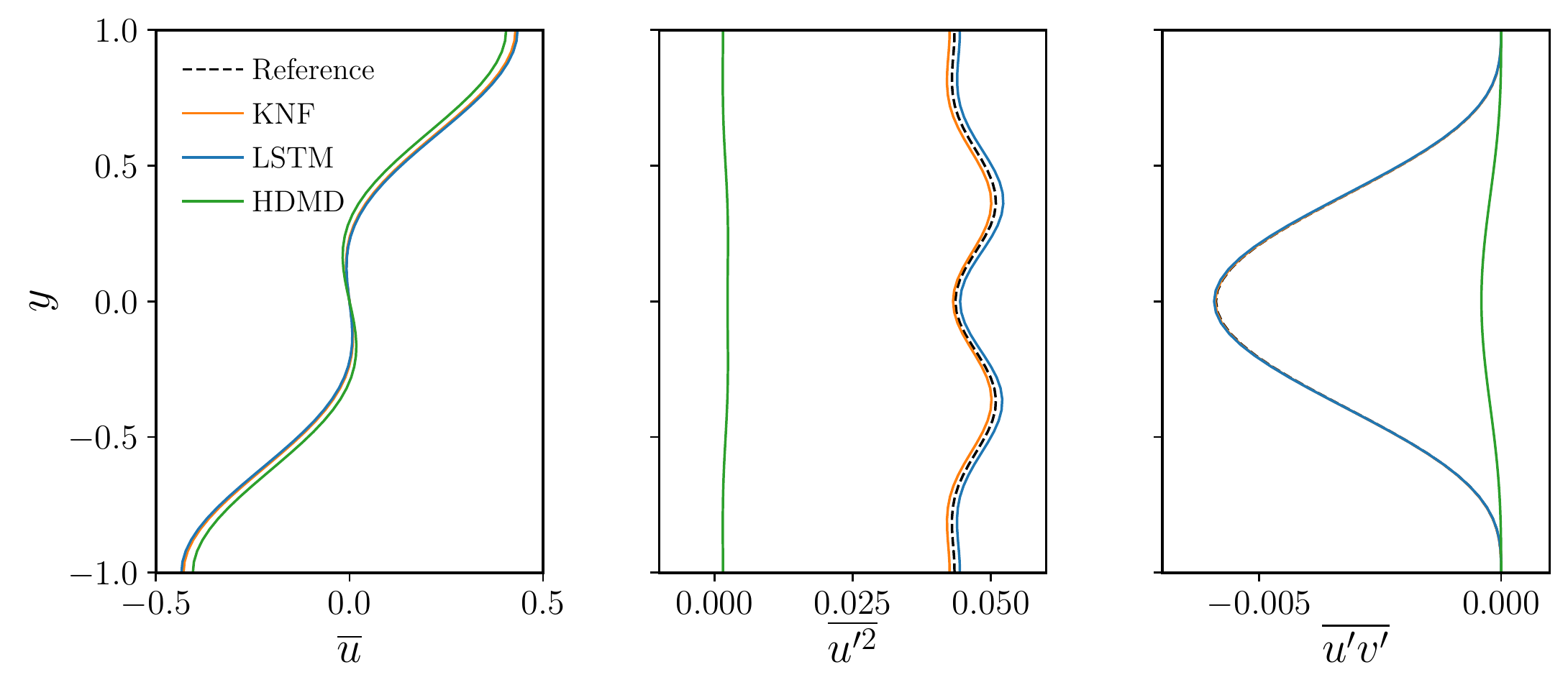}
    \caption{Long-term statistics corresponding to (left) streamwise mean profile, (middle) streamwise velocity fluctuations and (right) Reynolds shear stress. Results are presented for the reference nine-equation model \citep{moehlis_et_al}, an LSTM network with 1 layer and 90 neurons, trained with 10,000 datasets \citep{srinivasan}, and the KNF and HDMD models trained using a time series with $N$ = 10,000 and $q$ = 5.}
    \label{statistics}
\end{figure}

In our previous study~\citep{srinivasan}, we showed that using 10,000 time series for training, it is possible to obtain excellent reproduction of long-term statistics from the LSTM network, with $E_{\overline{u}} = 0.45\%$ and $E_{\overline{u^{\prime 2}}} = 2.49\%$. This was obtained with $p = 10$. In \Cref{statistics} we show a comparison of the long-term statistics obtained from the nine-equation model, the KNF and the HDMD methods, and this LSTM network, including mean flow, the streamwise fluctuations and the Reynolds shear-stress profile $\overline{u^{\prime}v^{\prime}}$. The KNF results are obtained from the KNF model with a performance close to the mean performance of the five KNF models, with $E_{\overline{u}} = 0.49\%$ and $E_{\overline{u^{\prime 2}}} = 1.54\%$ (see Table~\ref{KNF_tst}). These results highlight the excellent predicting capabilities of the LSTM network, given that sufficient training data is provided, and of the KNF method, due to the ability of the network and the Koopman-based framework with nonlinear forcing to exploit the sequential nature of the data. In contrast, the linear HDMD model is not able to reproduce the long-term statistics due to the gradual decay of the predictions to a constant value, leading to the errors of $E_{\overline{u}} = 4.79\%$ and $E_{\overline{u^{\prime 2}}} = 86.88\%$.

\begin{figure}
    \centering
    \includegraphics[width=4in]{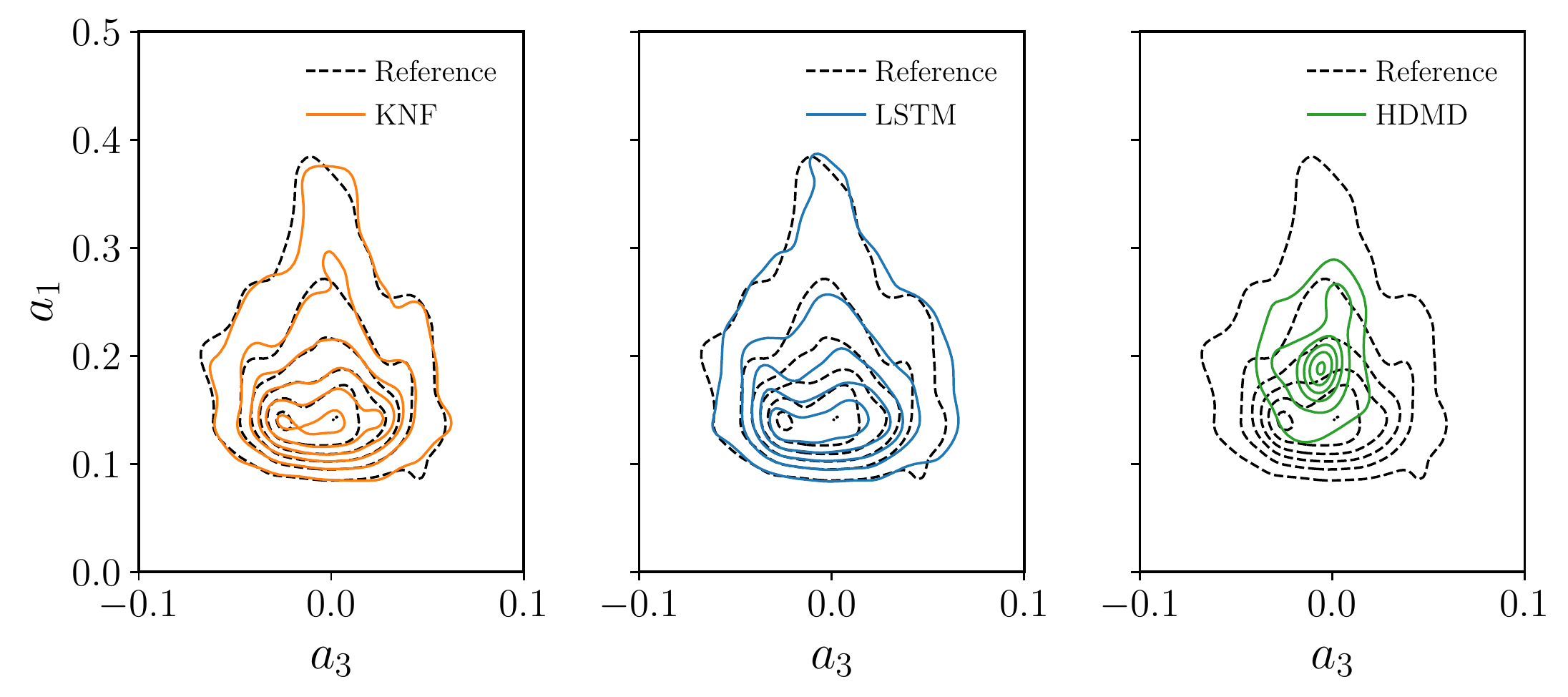}
    \includegraphics[width=4in]{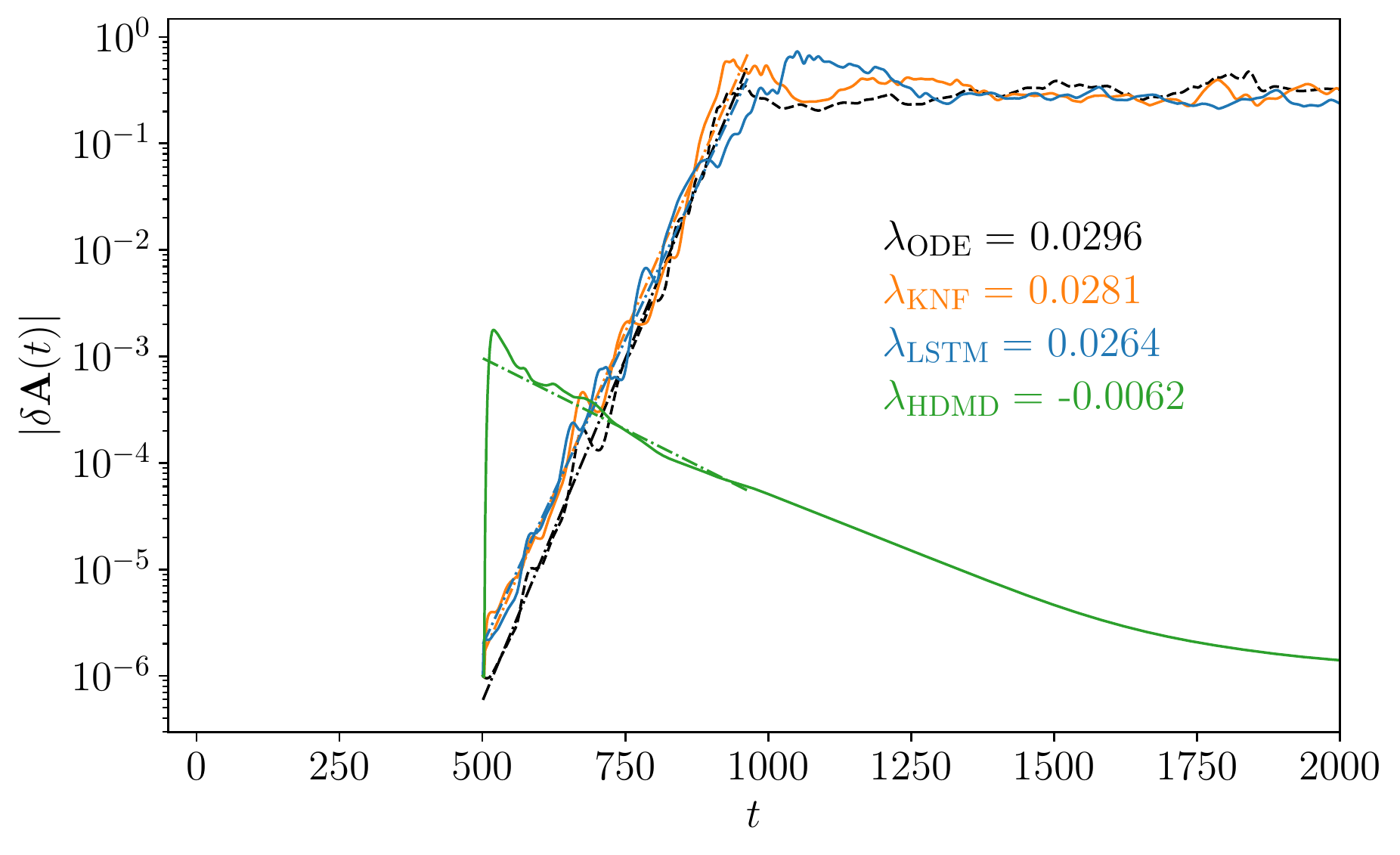}
    \caption{(Top three panels) Probability density function of the Poincar\'e maps, where the intersection with the $a_{2}=0$ plane (with ${\rm d}a_{2}/{\rm d}t <0$) is shown. (Bottom) Ensemble-averaged divergence of instantaneous time series after a perturbation with $| \delta \mathbf{A}_{0} |=10^{-6}$ is introduced at $t_{0}=500$, showing initial exponential growth and the value of the Lyapunov exponent (dashed lines added to illustrate the obtained slope). Results are depicted for LSTM, KNF, and HDMD models, comparing to the results from the reference nine-equation model \citep{moehlis_et_al}.}
    \label{fig_dynamics2}
\end{figure}

The quality of the reproductions was further assessed in terms of the dynamic behavior of the system, first through the Poincar\'e map defined as the intersection of the flow state with the hyperplane $a_{2}=0$ on the $a_{1}-a_{3}$ space (subjected to ${\rm d}a_{2}/{\rm d}t <0$). This map essentially shows the correlation between the amplitudes of the first and third modes, {\it i.e.} the modes representing the laminar profile and the streamwise vortices in the nine-equation model. In \Cref{fig_dynamics2}~(top three panels) we show the probability density function (pdf) of the Poincar\'e maps constructed from the 500 time series obtained from the KNF, LSTM, and HDMD predictions and the reference nine-equation model. In these figures, it can be observed that the LSTM network and the KNF method capture the correlation between the amplitudes of both modes, which indicates that their interaction is adequately represented by the NN and the Koopman-based framework with nonlinear forcing. However, the HDMD method does not reproduce this correlation adequately. Moreover, the KNF results are in a closer agreement with those from the nine-equation model. We also studied the separation among trajectories obtained from the reference model, the LSTM network and the KNF and HDMD methods by means of Lyapunov exponents. For two time series 1 and 2, we define the separation of these trajectories as the Euclidean norm in nine-dimensional space: 
 \begin{equation}
\left | \delta \mathbf{A}(t) \right | = \left [ \sum_{i=1}^{9} \left (a_{i,1}(t)-a_{i,2}(t)  \right )^{2} \right ]^{1/2},
\end{equation} 
and denote the separation at $t=t_{0}$ as $ | \delta \mathbf{A}_{0} |$. The initial divergence of both trajectories can be assumed to behave as: $\left | \delta \mathbf{A}(t') \right | = \exp (\lambda t') \left | \delta \mathbf{A}_{0} \right |$, where $\lambda$ is the so-called Lyapunov exponent and $t'=t-t_{0}$. We introduced a perturbation with norm $| \delta \mathbf{A}_{0} | =10^{-6}$ (which approximately corresponds to the accuracy of the current LSTM architecture \citep{srinivasan}) at $t_{0}=500$, where all the coefficients are perturbed, and then we analyzed its divergence with respect to the unperturbed trajectory. 
In \Cref{fig_dynamics2}~(bottom) we show the evolution of $ | \delta \mathbf{A}(t) |$ with time for the reference, as well as the LSTM, KNF, and HDMD predictions, after ensemble averaging 10 time series. All the three rates of divergence for the reference, LSTM, and KNF models are very similar, with almost identical estimations of the Lyapunov exponents $\lambda$: 0.0264 for the LSTM, 0.0281 for the KNF, and 0.0296 for the nine-equation model. Also note that after around approximately 500 time units of divergence, all three curves saturate. However, the HDMD model failed to capture the chaotic nature of the system, leading to a negative value for $\lambda$. This result provides additional evidence supporting the excellent reproductions of the dynamic behavior from the original system when using the present LSTM architecture and the Koopman-based framework with nonlinear forcing.

The excellent performance of the KNF method is interesting since it needs a much smaller data set, namely $0.025\%$ of that from the LSTM, for the training of the model. Moreover, the mapping matrices of $\mat{A}$ and $\mat{B}$ are computed in one shot, thus the KNF algorithm is orders of magnitude faster than the backpropagation algorithm, which is used for training the LSTM network, overall the required training time for the LSTM network is around four orders of magnitude larger than that of the KNF method. Note that the training time for KNF models increase linearly with the length of the training sequences, whereas a constant number of updates (\textit{i.e.} a smaller number of epochs) is sufficient to reach similar levels of accuracy. Our results show that the training time of the KNF method is comparable to that of the echo state network (ESN)~\citep{Pandey_esn}, while both the KNF method and LSTM network outperform the ESN in the reproduction of long-term statistics. 

\begin{figure}[h]
    \centering
    \includegraphics[width=3in]{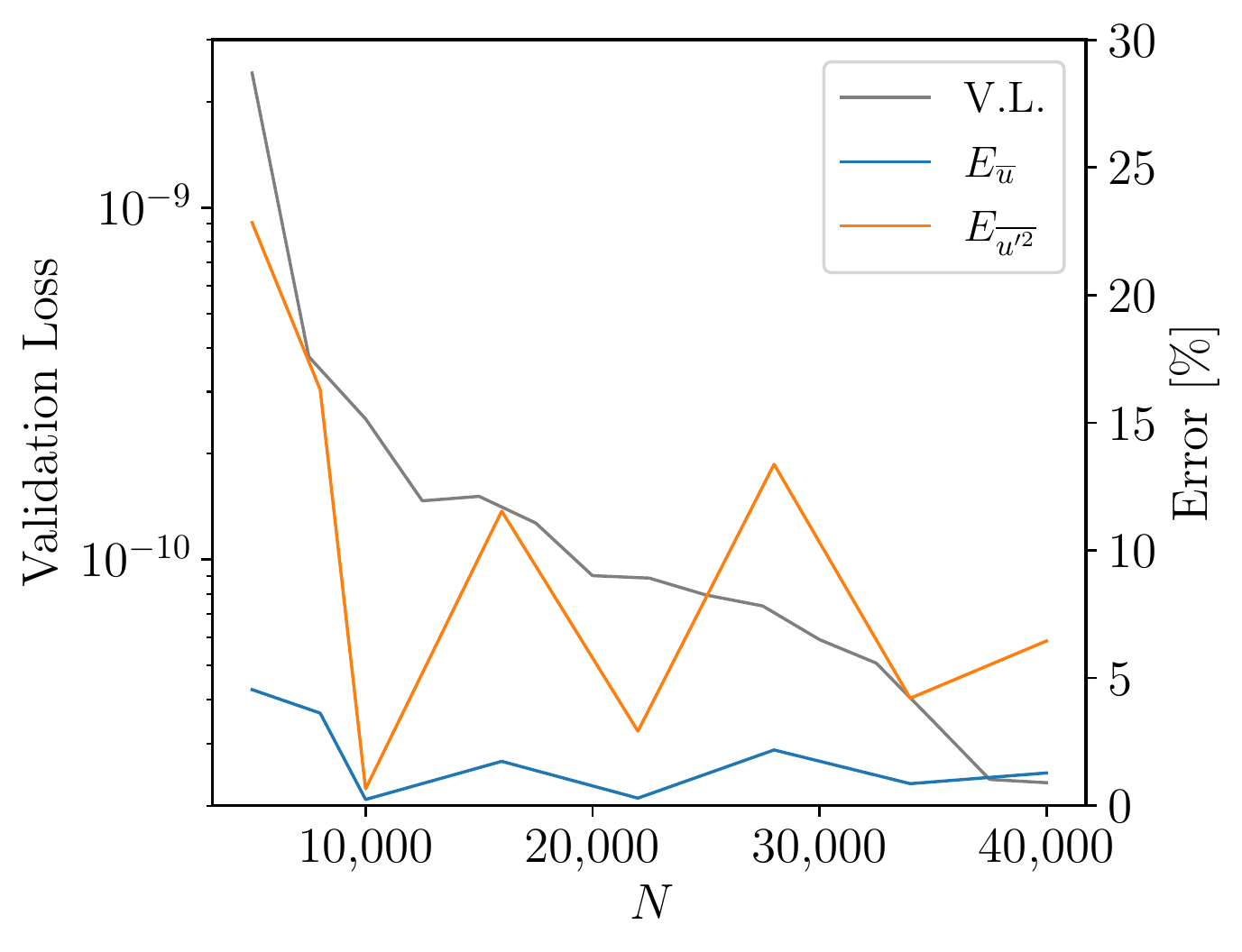}
    \caption{Validation loss and statistical errors versus the number of time units of the training time series for the KNF method with a delay dimension of 5.}
    \label{err_N_KNF}
\end{figure}

We have shown in \Cref{fig:knf_table_data} that increasing the amount of training data for the KNF method will not necessarily lead to improved reproductions of the long-term statistics. Such an increase of training data can be achieved either by increasing the number of time units of the training data set $N$ or increasing the dimension of the delay embedding $q$. It is also of interest to evaluate the effect of more training on the instantaneous predictions. With this purpose, the loss function of the NNs is utilized to represent the error on one step predictions of the validation data set, so it is possible to have a comparative understanding about the performance of the KNF method against the LSTM network. Note that the validation loss for both the LSTM network and KNF method is the sum of errors over the predicted sequence.~\Cref{err_N_KNF} shows the validation loss and statistical errors $E_{\overline{u}}$ and $E_{\overline{u^{\prime 2}}}$ versus the number of time units of the training data set $N$, for the KNF model with a delay dimension of 5. Here, we considered a time series with 40,000 time units as the validation set for the KNF method. For $N < 10,000$, the error in the instantaneous predictions and the statistical quantities both decrease with an increasing size of the training data set. Moreover, it can be seen that the error in instantaneous predictions is reduced with a further increase of $N$, a fact that indicates an improvement of instantaneous predictions with increasing amount of training data. However, this figure shows that better instantaneous predictions do not necessarily lead to a better approximation of the long-term statistics. For $N > 10,000$, utilizing more data for training leads to an increase of the error in the statistics while the instantaneous error still follows a decreasing trend. As discussed in $\S$\ref{sec_NNs_improve}, a similar behavior is observed for the LSTM, a fact that can be used to define a model-selection criterion for training.

\section{Robustness of the KNF framework}
\label{sec_KNF_robustness}

In this section, we discuss the capability of the KNF framework to reproduce the dynamics of the nine-equation model without a priori knowledge of the nonlinearities in the system and in the presence of noise. As discussed in $\S$\ref{sec-introduction}, the nine-equation model of \cite{moehlis_et_al} comprises nine ODEs, each of them involving a linear term and several nonlinear terms, all in the form of quadratic nonlinearities ($a_{i}a_{j}$), and they may include a constant. By accurate selection of the form of nonlinearities and construction of the forcing term, the KNF model given by \cref{Eq_m2} can closely resemble \cref{Eq:dyna_sys}, which may represent many dynamical systems of interest.

\subsection{Unknown nonlinearities}

We aim to construct the KNF model in a fully data-driven fashion and without consideration of information on nonlinearities. In particular, we construct the forcing term $\mathbf{f}$ as a library of candidate nonlinear functions and let the model identify the most important forms of nonlinearities from the data. We perform four tests by involving any possible quadratic, cubic, and quartic nonlinear terms in the KNF models, {\it i.e.} $(\mathbf{x})^{p_{2}}$, $(\mathbf{x})^{p_{3}}$, and $(\mathbf{x})^{p_{4}}$, to assess their performance when the types of nonlinearities in the dynamics are unknown. In the last test, we exclude the quadratic terms, which are the actual forms of nonlinearities in the original data, and include cubic and quartic nonlinear terms. In all tests, we also consider a constant term in $\mathbf{f}$. The KNF models are trained using a time series containing 10,000 time steps with the $\Delta t = 1$ and a delay-embedding dimension $q$ equal to 5. To generate a reference test set, we solve the ODEs of the nine-equation model, starting from 500 randomly generated initial conditions, each for a span of 4,000 time units.

\Cref{tab:KNF_tests} presents the number of active terms for different forms of nonlinearities in the KNF models, which are identified by the model using a threshold value of $\varepsilon = 0.05$. The total number of possible forms of quadratic, cubic, and quartic nonlinearities are 45, 165, and 495. The nine-equation model contains 30 different forms of quadratic nonlinearities. It can be seen that for the KNF--2 model 34 forms of quadratic nonlinearities are active out of 45 possible forms. In fact, the 30 quadratic terms of the nine-equation model are identified correctly, and there are 4 extra terms. For the KNF--23 and KNF--234 models, one extra quadratic term and only few cubic and quartic terms are identified as active. For KNF--34, which excludes quadratic terms, the model includes 79 and 31 cubic and quartic terms, respectively.
\begin{table}[h]
    \centering
    \caption{Number of active terms for different forms of nonlinearities identified by the selection algorithm (\Cref{algo:ridge}) from the training data, which are used in the KNF models. The numbers in brackets denote the maximum number of nonlinear terms of that order.}
    \resizebox{\columnwidth}{!}{%
    \begin{tabular}{ccccc}
        \hline \hline
        \rule{0pt}{3ex} Model & Included nonlinear terms & $(\mathbf{x})^{p_{2}}$ (45) & $(\mathbf{x})^{p_{3}}$ (165) & $(\mathbf{x})^{p_{4}}$ (495) \\
        \vspace{-0.3cm} \\
        \hline
        \vspace{-0.3cm} \\
        KNF--2 & quadratic   &   34 & -- & -- \\
        KNF--23 & quadratic, cubic   &   31 & 19 & -- \\
        KNF--234 & quadratic, cubic, quartic   &    31 & 19 & 4 \\
        KNF--34 & cubic, quartic  &   --  & 79 & 31 \\
        \vspace{-0.3cm} \\
    \hline \hline
    \end{tabular}
    }
    \label{tab:KNF_tests}
\end{table}

\begin{figure}
    \centering
    \includegraphics[width=4in]{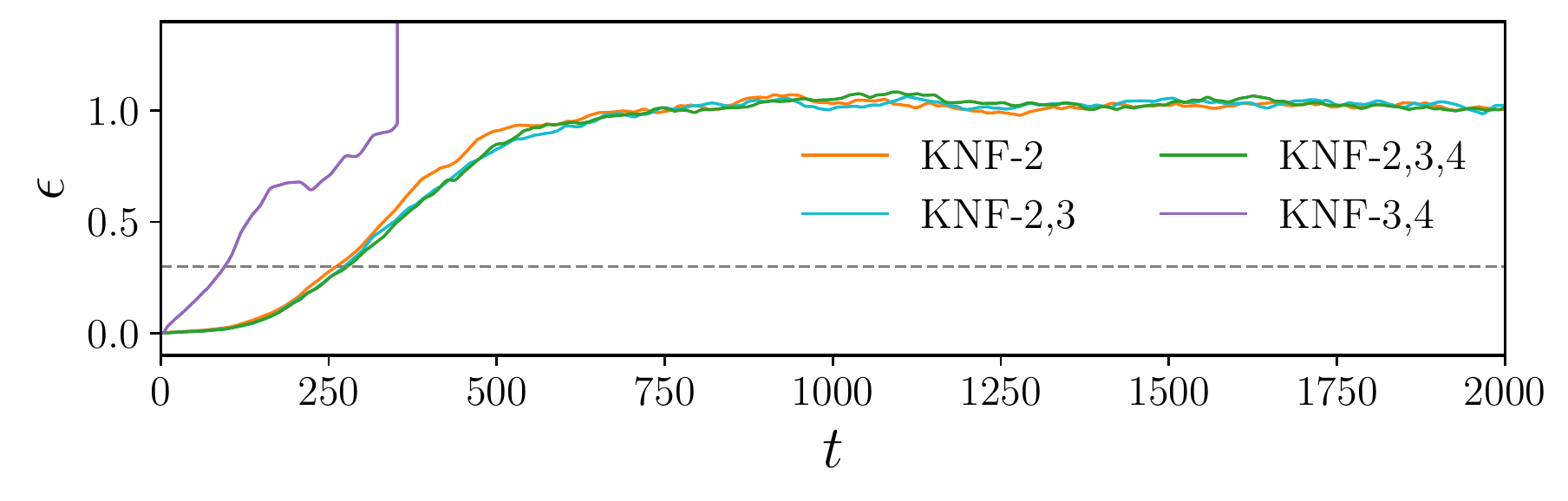}
    \caption{Relative Euclidean norm of errors $\epsilon(t)$ averaged over 500 randomly chosen initial conditions, as defined in \cref{Eq_err_short_term} for KNF models containing different forms of nonlinearities according to \Cref{tab:KNF_tests}. The dashed horizontal line shows the threshold value considered for accurate predictions, namely $\epsilon=0.3$.}
    \label{fig:Err_knfs}
\end{figure}

After training, all the time series of the reference test set are predicted using each model. \Cref{fig:Err_knfs} depicts $\epsilon(t)$ for different KNF models. Results show that all the KNF models containing quadratic nonlinear terms provide very good short-term predictions up to $t \simeq 280$ in average for 500 time series of the test set. For the KNF--34 model, in which the quadratic terms are excluded, the predictions diverge quickly.

The relative error in the reproduction of $\overline{u}$ and $\overline{u^{\prime 2}}$ are reported in \Cref{tab:statistics_knfs}. It can be observed that the KNF method can provide an excellent reproduction of the statistics with less than 1\% error. Moreover, the KNF models that involve higher-order polynomial nonlinearities can reproduce the statistics as well as the KNF--2 model. The KNF--34 model cannot reproduce the statistics due to the divergence of the time series. These results indicate that by a broad choice of entries in the forcing term $\mathbf{f}$ of the KNF method, the model can identify the most important forms of nonlinearities in the data and utilize them for accurate short-term prediction and reproduction of the long-term statistics.

\begin{table}[h]
    \centering
    \caption{Relative errors in the reproduction of the long-term statistics using KNF models containing different form of nonlinearities according to \Cref{tab:KNF_tests}.}
    \begin{tabular}{ccc}
        \hline \hline
        \rule{0pt}{3ex} Model & $E_{\overline{u}}[\%]$ & $E_{\overline{u^{\prime 2}}}[\%]$\\
        \vspace{-0.3cm} \\
        \hline
        \vspace{-0.3cm} \\
        KNF--2   &   0.28 & 0.57 \\
        KNF--23   &   0.12  & 0.50 \\
        KNF--234  &   0.44  & 0.77 \\
        KNF--34 &   --  & -- \\
        \vspace{-0.3cm} \\
    \hline \hline
    \end{tabular}
    \label{tab:statistics_knfs}
\end{table}

\subsection{Effect of noise}

In practice, the measurements of the system may be contaminated with noise, for instance due to experimental measurement inaccuracies. Thus, it is worthwhile to explicitly add noise to the data and examine the performance of the model. To this end, we add noise to the training data set as random entries with a normal (Gaussian) distribution with zero mean and a standard deviation of $\sigma_\mathrm{noise}$. We define noise percentage $\eta$ as $\sigma_{\mathrm{noise}}/\sigma_{\mathrm{data}} \times 100$, where $\sigma_{\mathrm{data}}$ is calculated for all the dimensions of all 500 time series. The reference test data set remains unchanged. We include all the possible quadratic, cubic, and quartic nonlinear terms in the forcing terms of the KNF models and train them on noisy data with $\eta$ values of 0.5\%, 1\%, 5\%, and 10\%. 

\begin{figure}
    \centering
    \includegraphics[width=4in]{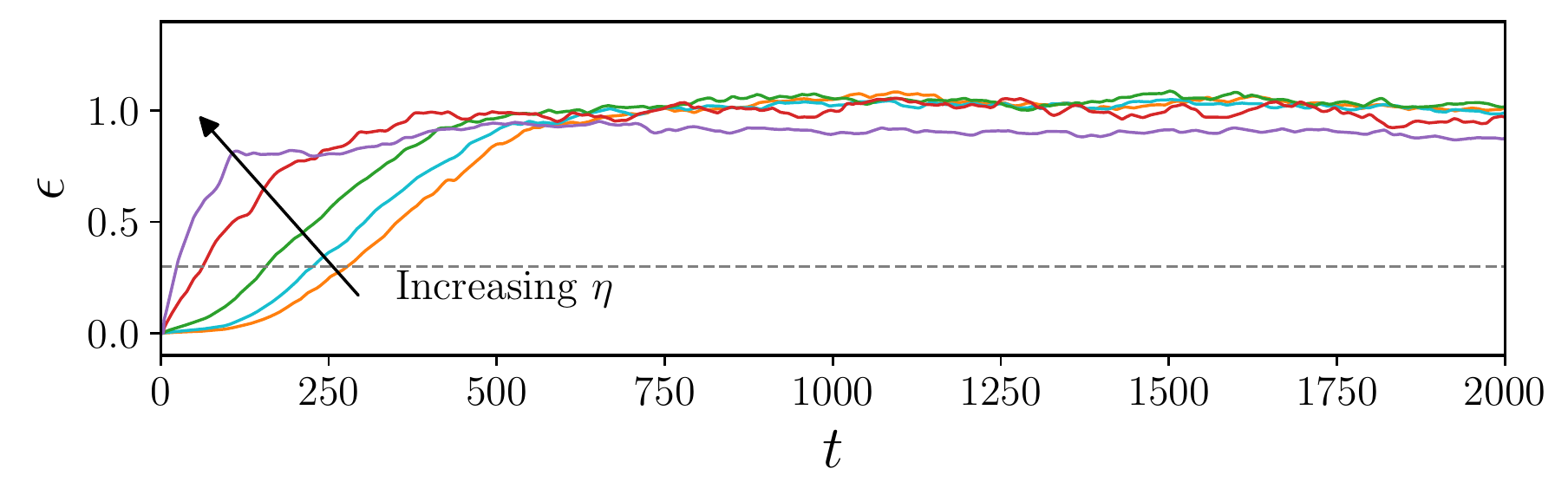}
    \caption{Relative Euclidean norm of errors $\epsilon(t)$ averaged over 500 randomly chosen initial conditions for KNF models trained on noisy data with $\eta$ of 0.5\%, 1\%, 5\%, and 10\%, as well as clean data. The dashed horizontal line shows the threshold value considered for accurate predictions, namely $\epsilon=0.3$.}
    \label{fig:Err_noise}
\end{figure}

The relative Euclidean norm of errors $\epsilon(t)$ for KNF models trained on noisy data are depicted in \Cref{fig:Err_noise}. These results are compared with those of the KNF model trained on clean data (KNF--234 from \Cref{tab:statistics_knfs}). It can be observed that although the error increases for large noise percentages, with $\eta=0.5\%$ and 1\% the KNF models can provide very good short-term predictions, exhibiting $\epsilon(t)< 0.3$ for 227 and 155 time units, respectively. These models can also provide an excellent reproduction of the long-term statistics with errors lower than 5\% as can be seen in \Cref{tab:noise}. Note that, as expected, higher noise percentage leads to an increase in the errors.

\begin{table}[h]
    \centering
    \caption{Errors in the reproduction of the long-term statistics obtained from the KNF models trained on noisy data. All the possible quadratic, cubic, and quartic nonlinearities are included in the KNF models.}
    \begin{tabular}{ccc}
        \hline \hline
        \rule{0pt}{3ex} $\eta [\%]$ & $E_{\overline{u}}[\%]$ & $E_{\overline{u^{\prime 2}}}[\%]$\\
        \vspace{-0.3cm} \\
        \hline
        \vspace{-0.3cm} \\
        Clean   &   0.44  & 0.77 \\
        0.5   &   0.54  & 2.32 \\
        1  &   1.36  & 4.39 \\
        5  &   5.06  & 24.54 \\
        10  &   13.52  & 53.13 \\
        \vspace{-0.3cm} \\
    \hline \hline
    \end{tabular}
    \label{tab:noise}
\end{table}

\Cref{tab:noise2} shows the number of active terms for different forms of nonlinearities. These terms are identified by the selection algorithm (\Cref{algo:ridge}) from the noisy data with a threshold $\varepsilon$ equal to 0.05 and then used in the KNF models. It can be seen that the iterative procedure for identification of the active terms (\Cref{algo:ridge}) is robust against noise where the same terms are identified as active from the noisy data for $\eta$ up to 1\%. For larger values of noise, the number of active terms is increased.

\begin{table}[h]
    \centering
    \caption{Number of active terms for different forms of nonlinearities identified by the selection algorithm (\Cref{algo:ridge}) from the training data contaminated with noise, where all the possible quadratic, cubic, and quartic nonlinearities are included. The numbers in brackets denote themaximum number of nonlinear terms of that order.}
    \begin{tabular}{cccc}
        \hline \hline
         \rule{0pt}{3ex} $\eta[\%]$ & $(\mathbf{x})^{p_{2}}$ (45) & $(\mathbf{x})^{p_{3}}$ (165) & $(\mathbf{x})^{p_{4}}$ (495) \\
        \vspace{-0.3cm} \\
        \hline
        \vspace{-0.3cm} \\
        clean   &   31 & 19 & 4 \\
        0.5   &   31 & 19 & 4 \\
        1    &    31 & 19 & 4 \\
        5   &   33  & 26 & 5 \\
        10  &   40  & 52 & 12 \\
        \vspace{-0.3cm} \\
    \hline \hline
    \end{tabular}
    \label{tab:noise2}
\end{table}

Our results show that in the case of nine-equation model of \cite{moehlis_et_al}, which is a low-order model of near-wall turbulence, the KNF method can provide an excellent reproduction of the dynamics in a fully data-driven fashion. For a high-dimensional chaotic flow, it was shown by \cite{khodkar2019} that the KNF method can obtain very good short-term predictions for a high-Reynolds 2D lid-driven cavity flow, where the Reynolds-stress terms at each sampled grid point, which are quadratic nonlinear terms, were used to form the nonlinear forcing vector $\mathbf{f}$. However, in their work, the performance of the model was not assessed in the reproduction of long-term dynamics and statistics. The present results suggest the possible applicability of the KNF method in learning the dynamics of turbulence; however, the success of  this approach would depend on the ability of the method to model order reduction (a topic which is not addressed in this work), as well as to represent nonlinear dynamics. These aspects should be investigated through a more general model assessment.

\section{Towards improving neural-network predictions}
\label{sec_NNs_improve}

The results in $\S$\ref{sec_comparison} showed that the LSTM network is able to accurately predict the temporal dynamics and statistics of a low-dimensional representation of near-wall turbulence. Next we explore different strategies to potentially improve the accuracy and efficiency of RNN predictions~\citep{guastoni_tsfp}.

\subsection{Validation loss and model-selection criterion}

\begin{figure}[t]
    \centering
    \includegraphics[width=3in]{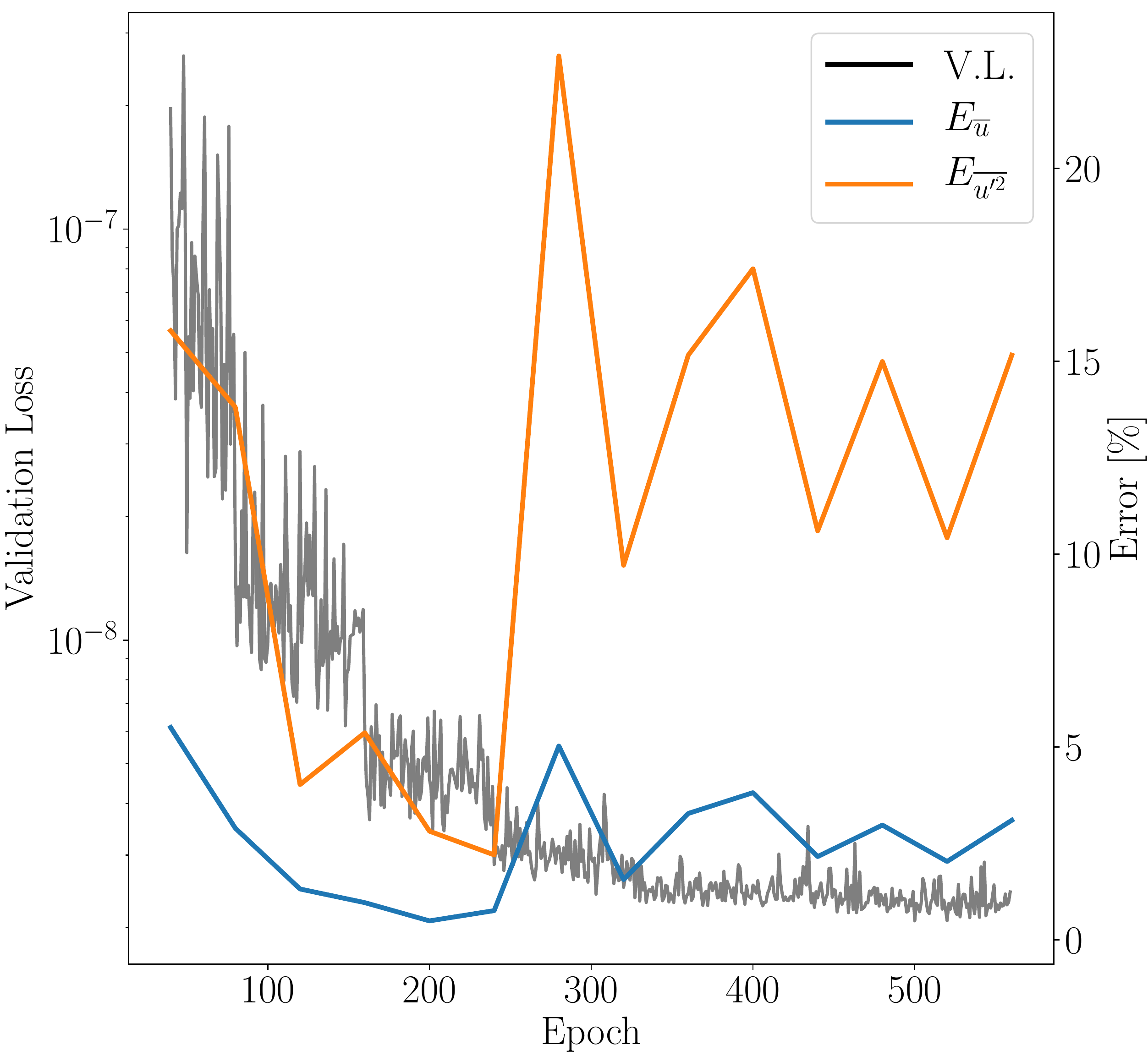}
    \caption{Evolution of the validation loss and the statistical errors as the training of the LSTM2--1--100 network progresses. }
    \label{err_vs_val}
\end{figure}

As discussed above, the amplitudes of the modes in the model by \cite{moehlis_et_al} exhibit fluctuations that are compatible with a chaotic turbulent state. Given the high sensitivity of the model to very small variations in the mode amplitudes, a loss function based on short-time horizon predictions, namely one time step ahead, is required to obtain satisfactory predictions. On the other hand the trained model needs to correctly reproduce not only the instantaneous behavior but also the statistical features of the original shear flow model. The approach used in the work by~\cite{srinivasan} involves a loss function based only on the error in the instantaneous prediction. Neural networks having at least one hidden layer have been shown to be universal approximators \citep{Cybenko}, hence they are in principle able to represent any real function. A perfect reproduction of the temporal behavior of the model would also provide correct mean and fluctuations at no added cost, however there is no guarantee that such a model can be learned and, even in that case, the model would theoretically be available after an infinitely long training. In order to verify to which extent the loss function based on instantaneous predictions represents an effective solution, different neural-network configurations were tested to assess the correlation between the achieved validation loss and the error in the statistics of the flow. In Table~\ref{table_lstm} we summarize the various LSTM architectures under study, where we vary the number of layers, the number of time series used for training and the time step between samples. The results from~\cite{srinivasan} are also reported for comparison. Let us consider the case LSTM2--1--100, consisting of 2 layers, with 90 units per layer, trained with 100 time series and a time step of 1. Figure~\ref{err_vs_val} shows the validation loss and the relative errors $E_{\overline{u}}$ and $E_{\overline{u^{\prime 2}}}$ for this network, as functions of the number of epochs trained ({\it i.e.}\ the number of complete passes through all the samples contained in the training set). In the initial stage of the training, starting from the randomized initialization of the weights and biases, the reduction of the error in the instantaneous behavior and in the statistical quantities show a similar trend. However, this figure also shows that (as in the case of the KNF method) lower validation loss values do not always lead to a better approximation of the long-term statistics. In fact, as the training progresses, the optimization algorithm continues to improve the short-term predictions, whereas beyond around 240 epochs the error in the statistics does not follow a descending trend anymore. The observed behavior is explained by the fact that the loss function does not contain any term explicitly related to the statistics which could guide the optimization algorithm towards parameter sets with a better representation of the statistical quantities. Note that since the initialization of the parameters of the network is random, the performance in the prediction of mean and fluctuation may vary when the same model is trained multiple times. The achievable accuracy and the epoch at which this value will be reached are unknown a priori.

In an effort to keep a simple loss function, we use the error in the statistics as criterion to halt the training when a minimum is reached for this value. Note that the error can vary significantly from one epoch to the other, hence it is advisable to consider multiple epochs to identify the general trend of the error curves. Doing so, we can achieve excellent predictions of the long-term statistics while using a simple loss function based on the instantaneous predictions of the coefficients. As shown in Table~\ref{table_lstm}, the improvement over the models reported in our previous work~\citep{srinivasan} is particularly evident for the models trained on the small dataset, yielding an accuracy in the statistics comparable with that of the networks trained with bigger data sets. It is also important to note that the improved scheduled reduction of the learning rate employed for the results in Table~\ref{table_lstm} allowed to obtain much lower validation losses than in our previous work, using a similar training time. When reaching such low values of the loss function, the trade-off between the instantaneous and the average performance is more apparent.

The outlined strategy is not the only one that can be implemented aiming to reduce the error on the statistics of the flow model, however such modifications to the training algorithm typically require additional hyper-parameters that need to be optimized in order to obtain a satisfactory performance. One possible approach consists in including a new term in the loss function accounting for the error in the long-term statistics. In this case the relative importance of the two terms needs to be adjusted, as prioritizing the accuracy of the statistics may lead to a model that learns only the average behavior of the system. Alternatively, it is possible to use the fact that the time horizon of the predictions influences which features of the problem are learnt by the neural network, as highlighted by \cite{Chiappa}. In that work it was shown how improvements in the short-term accuracy ({\it i.e.}\ in the prediction of the instantaneous behavior) come at the expense of the accuracy of global dynamics. Using the results of the network to make predictions several time steps ahead would encourage the network to learn the long-term behavior of the system and thus its global dynamics. As stated by \cite{Chiappa}, this approach has the added advantage of training the model in a way that is similar to its actual utilization. In fact, during the evaluation and usage, our networks rely only on the previous predictions after the first $p$ time steps. 

\begin{figure}
    \centering
    \begin{overpic}[width=0.79\textwidth]{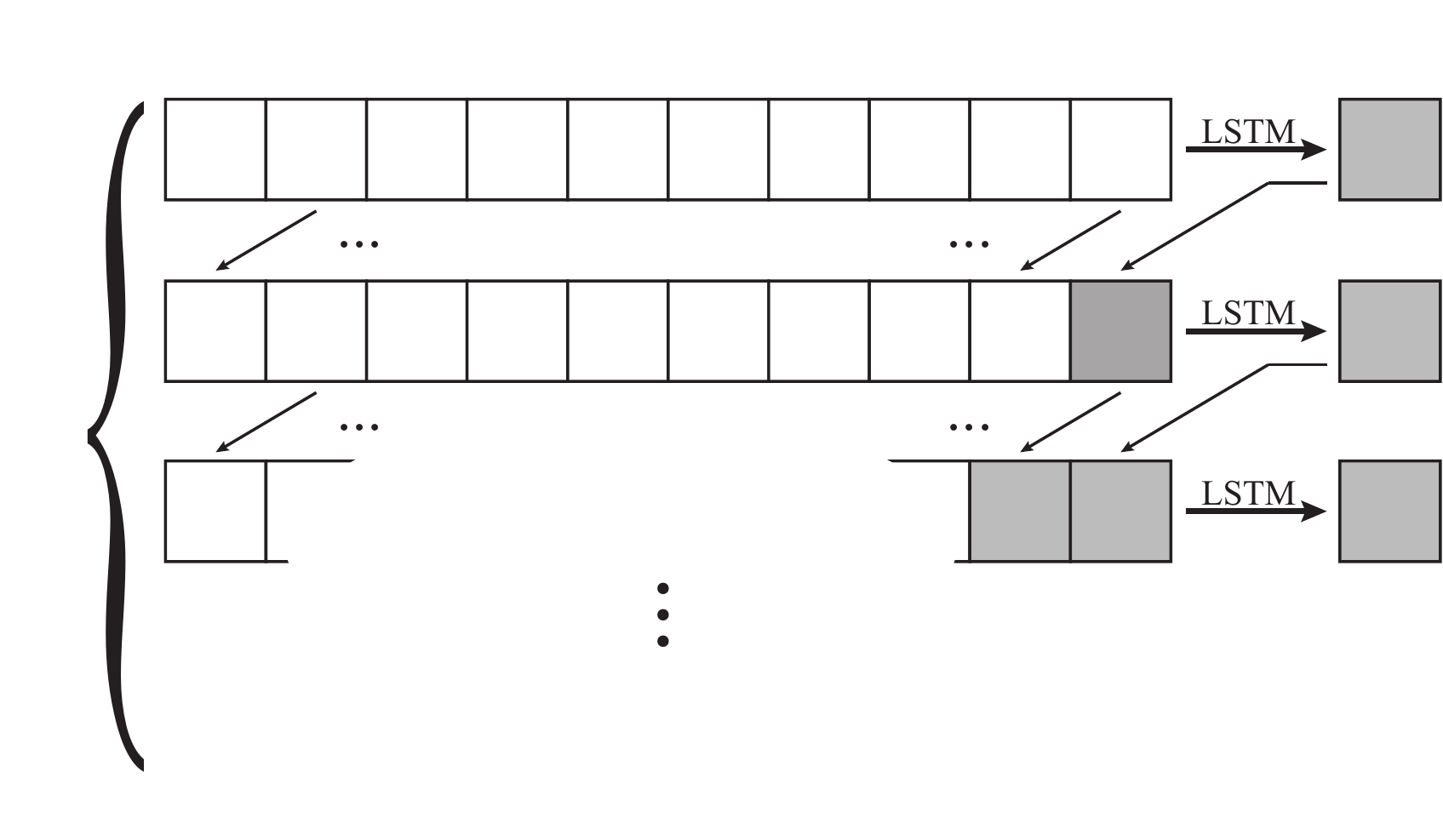}
    \put (1.5,27.5) {$n_p$}
    \put (12, 53) {$\mathbf{a}_{t-p}$}
    \put (75, 53) {$\mathbf{a}_{t-1}$}
    \put (67.5, 53) {$\mathbf{a}_{t-2}$}
    \put (95, 53) {$\mathbf{a}_{t}$}
    \end{overpic}
    \caption{Visual representation of the construction of the batch, when the LSTM model is trained using $n_p$ predictions at each parameter update.}
    \label{fig:pop}
\end{figure}
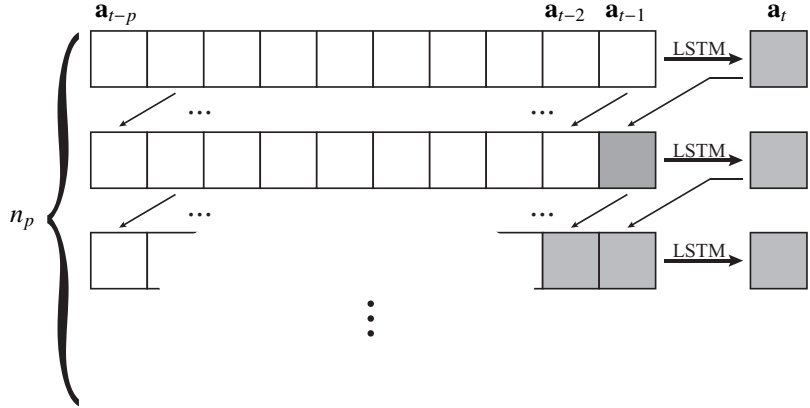

The latter approach requires to take into account the error in the current prediction based on previous predicted values, which results in a more complicated loss function which would require higher-order gradient descent for backpropagation. We can introduce a simplification, by considering each prediction as a separate sample, which enables an implementation with minimal modification to the original training. The construction of the batch of samples for the update algorithm, summarized in \Cref{fig:pop}, is performed as follows: let $n_p$ be the number of predictions that are considered at each update of the LSTM network, for each iteration $n_{\text{batch}}$ sequences of length $p$ are taken from the training dataset, then for each sequence the LSTM network is evaluated with the current set of network parameters in order to predict the $(p+1)^{\text{th}}$ values of the sequence. This prediction is concatenated with the previous $p-1$ values in the original time sequence and the result is used as input for the network to predict the $(p+2)^{\text{th}}$ values. This operation is repeated $n_p$ times, obtaining a batch of $n_p \times n_{\text{batch}}$ samples of length $p$, to be used for the gradient-descent algorithm.
An important downside of this implementation is the substantial increase in the computational cost of the training, in fact $n_p$ evaluations of the network need to be performed each update step to construct the batch. In particular, $n_p=8$ determines an increase of the computational time by a factor of 4 and this increases linearly with $n_p$ for higher values, as the network evaluations become the most computationally-expensive part of the training.
The training loss is averaged over the $n_p$ predictions, whereas the validation loss is defined as the MSE on the first prediction from ground-truth validation sequences, in order to compare the losses with previously-trained models. Similarly, inference is performed in the same way, independently from the number of predictions that are computed at each update step.

\begin{figure}
    \centering
    \includegraphics[width=0.89\textwidth]{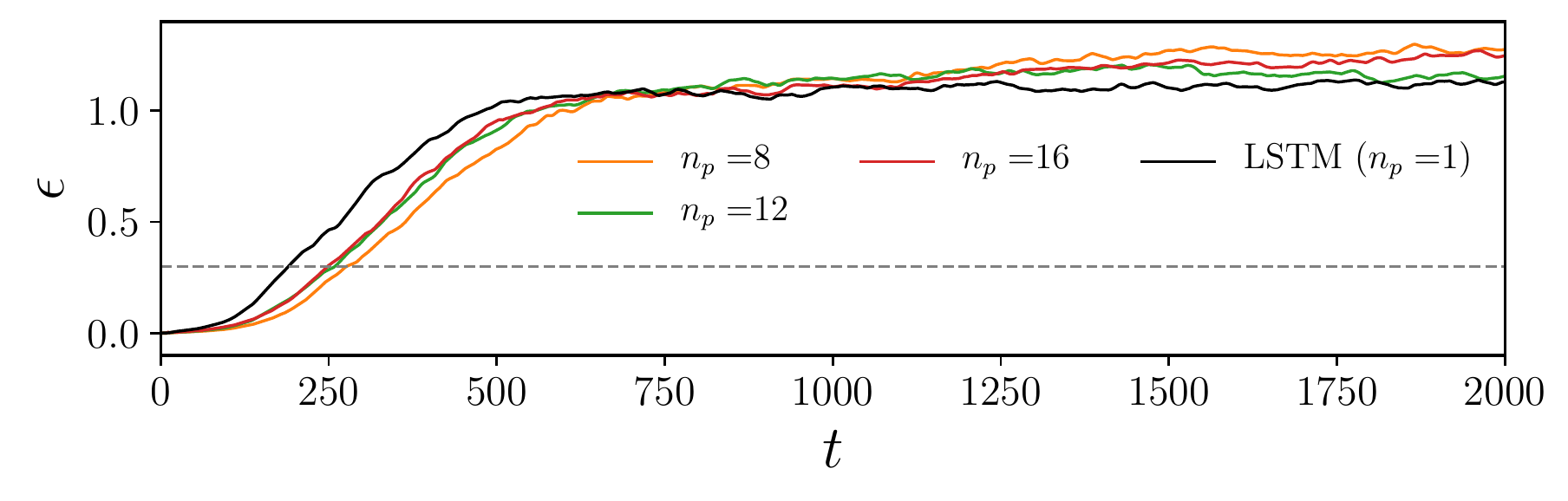}
    \caption{Relative Euclidean norm of errors $\epsilon(t)$ averaged over 500 randomly chosen initial conditions. The LSTM2--1--100 networks trained with a different number of successive predictions $n_p$ are compared. The dashed horizontal line shows the threshold value considered for accurate predictions, namely $\epsilon=0.3$.}
    \label{fig:eps_error_lstm_pop}
\end{figure}

The reference architecture is LSTM2--1--100, with 2 layers of 90 LSTM units each, trained on 100 time series with time step of 1. 
The results of the networks trained with $n_p=(8, 12, 16)$ are reported. Note that when one of the last two values is considered, there are samples in the batch that are completely made by predictions, without any value from the original time series. Because of the prediction error of the neural-network model, these can be categorized as ``noisy" samples. The successful training of these networks testifies the robustness of the LSTM architecture, further investigated in \S\ref{ssec_robustness_lstm}. 
The inclusion in the training of the predictions for a relatively low number of timesteps ($8<n_p<16$) can effectively modify the network performance on a much longer time, improving the accuracy of the predictions over an interval in the order of hundreds of timesteps, as shown by the relative Euclidean norm of errors $\epsilon(t)$ in \Cref{fig:eps_error_lstm_pop}. With the new training procedure the LSTM models can provide a short-term accuracy that is comparable with the KNF models reported in \Cref{Error-short-term}. Note however that the $\epsilon(t)$ stabilizes to a higher value when $n_p>1$. Despite the longer prediction window, the error in the statistics is found to be slightly higher than the $n_p=1$ counterpart and also in this case it does not follow a monotonically-descending trend, as in \Cref{err_vs_val}. The latter observation suggests that the mean and fluctuations errors should be explicitly included in the loss function when the statistical performance of the model is particularly relevant for its application.

\subsection{Effect of the time step}
The sequences provided to the neural network for training are evenly spaced in time, however the choice of the proper time step between data points $\Delta t$ depends on the problem at hand. The time step acts as a low-pass filter on the data, preventing the model from learning higher-frequency dynamics. On the other hand, for a fixed amount of samples, a larger $\Delta t$ allows to train the model over a longer time span. As shown in Table~\ref{table_lstm}, we considered the LSTM network with 1 layer and 90 neurons, and trained it using the same time series with $\Delta t=10$, 1 and 0.1 time units. Note that the input dimension is maintained constant by setting $p=10$. The number of time series and epochs for training were chosen so that it could be possible to compare models that have been trained on a similar number of samples. The results in Table~\ref{table_lstm} show that increasing the time step from 1 to 10 leads to a validation loss three orders of magnitude larger, a fact that indicates the difficulty in learning the model dynamics when such a coarse sampling in time is considered. On the other hand, reducing the time step from 1 to 0.1 does not yield any additional improvement in the predictions. The loss function has a similar trend and the final values are comparable when using time steps equal to 1 and 0.1, showing that most of the characteristics of the system have been properly captured. It may be possible to find a $\Delta t$ that further reduces the error based on the temporal characteristics of the signal. 
\begin{table*}[ht]
 \centering
 \caption{Summary of LSTM cases and their performance using different numbers of training data sets and time resolutions. The results marked with * are reported from~\citep{srinivasan} for comparison. Note that we employed 90 units and $p=10$ in all the cases. The statistical errors for LSTM1--10--1000 are not reported because the predictions exhibited a clearly non-physical behavior during all stages of training.}
 \label{table_lstm}
 \resizebox{\columnwidth}{!}{%
 \begin{tabular}{c c c c c c c}
 \hline
 \hline \vspace{-4pt}\\
 \rule[-1.6ex]{0pt}{0pt} Case & N. Layers & $\Delta t$ & Training data sets & $E_{\overline{u}}$ $[\%]$ & $E_{\overline{u^{\prime 2}}}$ $[\%]$ & Validation Loss\\ 
 \hline \\
LSTM1--1--100*   & 1 & 1 & 100    & 2.36 & 14.73 & $2.0 \times 10^{-8}$ \\
LSTM1--1--1000*  & 1 & 1 & 1,000  & 0.83 & 3.44  & $8.5 \times 10^{-9}$ \\
LSTM1--1--10000* & 1 & 1 & 10,000 & 0.45 & 2.49  & $5.2 \times 10^{-9}$ \\
LSTM2--1--100*   & 2 & 1 & 100    & 1.94 & 6.82  & $2.4 \times 10^{-8}$ \vspace{6pt}\\
 \hline \\
LSTM1--1--100 & 1 & 1 & 100 & 0.26 & 0.59 & $ 6.68 \times 10^{-9}$ \\
LSTM1--01--100 & 1 & 0.1 & 100 & 1.81 & 6.03 & $ 9.13 \times 10^{-10}$ \\
LSTM1--10--1000 & 1 & 10 & 1,000 & -- & -- &  $ 3.65 \times 10^{-5}$ \\
LSTM1--1--1000 & 1 & 1 & 1,000 & 0.57 & 0.58 & $ 8.36 \times 10^{-9}$ \\
LSTM1--01--1000 & 1 & 0.1 & 1,000 & 1.18 & 1.39 & $ 6.46 \times 10^{-9}$ \\
LSTM1--1--10000 & 1 & 1 & 10,000 & 0.31 & 0.48 &  $ 9.85 \times 10^{-9}$ \\
LSTM2--1--100 & 2 & 1 & 100 & 0.80 & 1.13 & $ 8.39 \times 10^{-9}$ \\
LSTM2--1--1000 & 2 & 1 & 1,000 & 0.54 & 0.62 & $ 8.84 \times 10^{-9}$ \\
LSTM2--1--10000 & 2 & 1 & 1,000 & 0.69 & 1.37 & $ 2.72 \times 10^{-9}$ \vspace{6pt}\\
 \hline
 \hline
 \end{tabular}
    }
 \end{table*}

\subsection{Robustness of the LSTM models}\label{ssec_robustness_lstm}

Here we investigate robustness in the performance of the LSTM network, as we did for the KNF method in $\S$\ref{sec_comparison} and $\S$\ref{sec_KNF_robustness}. First, we employ the LSTM1--1--100 architecture, consisting of 1 layer with 90 neurons, trained with 100 time series and a time step of 1. The model is trained with a simple loss function, where we use 20\% of the training data as a validation set and consider early stopping to avoid overfitting. We evaluate the performance of the model in the reproduction of the statistics for four different reference test sets, each containing 500 time series. As before, each of these test sets is created from 500 randomly generated initial conditions, and each time series is produced for 4,000 time units. Results show that the performance of the LSTM model is robust for different reference test sets, where standard deviations in the reproduction of the statistics are $0.13\%$ and $1.67\%$ for $\sigma(E_{\overline{u}})$ and $\sigma(E_{\overline{u^{\prime 2}}})$, respectively, over all tests.

\begin{figure}
    \centering
    \includegraphics[width=4in]{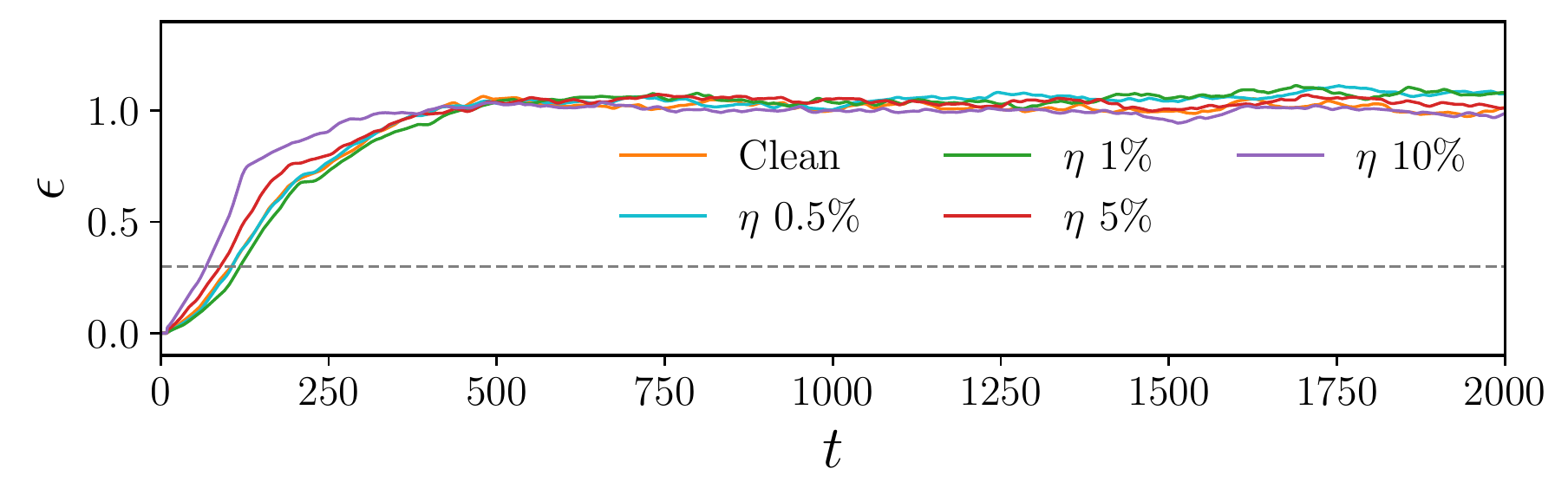}
    \caption{Relative Euclidean norm of errors $\epsilon(t)$ averaged over 500 randomly chosen initial conditions for LSTM models trained on noisy data with $\eta$ of 0.5\%, 1\%, 5\%, and 10\% and clean data. The dashed horizontal line shows the threshold value considered for accurate predictions, namely $\epsilon=0.3$.}
    \label{fig:Err_noise_lstm}
\end{figure}

In the next step, we investigate the effect of noise on the performance of the LSTM network. We utilize the same architecture (LSTM1--1--100) and train the network on noisy data sets with noise levels $\eta$ of 0.5\%, 1\%, 5\% and 10\%, and then evaluate the performance of the network on a clean reference test set containing 500 time series. The results for the short-term predictions are depicted in \Cref{fig:Err_noise_lstm} as the relative Euclidean norm of errors $\epsilon(t)$ averaged over 500 time series with randomly chosen initial conditions. Moreover, we compute the errors in the reproduction of the statistics for each model and calculate the standard deviation of the results for all the models as $1.27\%$ for $\sigma(E_{\overline{u}})$ and $7.52\%$ for $\sigma(E_{\overline{u^{\prime 2}}})$. The LSTM model trained on noisy data with $\eta=10\%$ leads to relative errors of $E_{\overline{u}} = 4.24\%$ and $E_{\overline{u^{\prime 2}}} = 17.59\%$. These results show that both the short-term predictions and the reproduction of long-term statistics are robust against noise for the LSTM network. In this case, the LSTM network performs better than the KNF method, where even the LSTM model trained on a noisy data set with $\eta=10\%$ provides an acceptable reproduction of the long-term statistics.

\subsection{Use of gated recurrent units (GRUs)}
The performance of an alternative type of RNN, the so-called gated recurrent unit (GRU), is also studied here. The structure of GRU layers is simpler than in the LSTM, consisting of a single \textit{update gate} instead of the forget and input gates. Also, the cell state and the output are merged into a single vector. The network architecture considered here has 1 layer of 90 nodes and it is similar in every aspect to the corresponding LSTM case, except for the node definition. The number of parameters that need to be optimized is smaller than in the LSTM, and therefore GRUs should require less computational resources to be trained. In our experience however, when training the considered architecture on CPU, the LSTM network was approximately as fast as its GRU counterpart. Despite the fact that it is possible to obtain similar validation losses with GRUs and LSTM networks, the resulting errors in the statistics are significantly higher in the former. In particular, when training with only 100 time series the predicted results exhibited a non-physical behavior. Although the results in Table~\ref{table_gru} suggest that the predictions may improve when using much larger training databases, the LSTM networks provide much more accurate predictions and they are therefore preferred for the present application.

\begin{table*}[ht]
 \centering
 \caption{Summary of GRU cases and their performance using different numbers of training data sets. Note that in all the cases 1 layer of 90 units was employed, with $p=10$. The statistical errors for GRU100 are not reported because the predictions exhibited a clearly non-physical behavior during all stages of training.}
 \label{table_gru}
 \begin{tabular}{c c c c c}
 \hline
 \hline \vspace{-4pt}\\
 \rule[-1.6ex]{0pt}{0pt} Case & Training data sets & $E_{\overline{u}}$ $[\%]$ & $E_{\overline{u^{\prime 2}}}$ $[\%]$ & Validation Loss\\ 
 \hline \vspace{-4pt}\\ 
GRU100 & 100 & -- & -- & $1.33 \times 10^{-8}$ \\
GRU1000 & 1,000 & 2.30 & 12.49 & $6.13 \times 10^{-9}$ \\
GRU10000 & 10,000 & 3.05 & 2.61 & $5.61 \times 10^{-9}$ \vspace{6pt} \\
 \hline
 \hline
 \end{tabular}
 \end{table*}

\section{Summary and conclusions}
\label{sec_conclusions}
In this study we assessed the feasibility of using RNNs and Koopman-based frameworks to predict the temporal dynamics of the low-order model of near-wall turbulence by \cite{moehlis_et_al}. Our previous results \citep{srinivasan} indicated that it is possible to obtain excellent reproductions of the long-term statistics using LSTM networks. Here we show that it is possible to obtain the same level of accuracy for long-term predictions by utilizing the Koopman framework with nonlinearities modeled through external forcing. Both approaches are able to reproduce the temporal dynamics of the system characterized through \emph{e.g.}\ Poincar\'e maps and Lyapunov exponents. However, the KNF method requires much less data and time for training: a data set with the size of $0.025\%$ of that from the LSTM is sufficient to train the KNF model. When considering the smallest datasets used (1 sequence of 10,000 steps for the KNF model and 100 sequences of 4000 steps for the LSTM), the training time of the LSTM network is about four orders of magnitude larger than that of the KNF model. Our results also indicate that the KNF method provides a longer prediction horizon for short-term forecasting in comparison with the LSTM network, producing accurate predictions (averaging over 500 time series) for 280 time units against 130 time units from a single-layer LSTM network, better results can be achieved adding a second LSTM layer, but at added computational cost. Our experiments suggest that KNF or other elaborate model designs can potentially outperform general-purpose machine-learning models such as LSTM. However, it should be considered that taking the nine-equation model as the reference gives an advantage to the KNF model over the LSTM network in the representation of the nonlinearities which is not present in real wall-bounded turbulence. The robustness of the models to noisy data has been tested, showing that the KNF model can provide acceptable results with a noise percentage $\eta$ up to 5\%. The LSTM network performs slightly better in this regard, maintaining a satisfactory performance with a noise percentage $\eta=10\%$.
Furthermore, we show that even using relatively small LSTM networks trained with low numbers of time series, {\it e.g.}\ the LSTM1--1--100 case, it is possible to obtain very low errors in the mean and the fluctuations, {\it i.e.}\ $E_{\overline{u}}=0.26\%$ and $E_{\overline{u^{\prime 2}}}=0.59\%$. It is important to highlight that a loss function based only on the instantaneous predictions of the mode amplitudes may not lead to the best predictions in terms of statistics, and it is necessary to define a model-selection criterion based on the values of $E_{\overline{u}}$ and $E_{\overline{u^{\prime 2}}}$. 
A training procedure for the LSTM network that included in the loss function the error on predictions based on previous predictions has been implemented. At added computational cost, it is possible to achieve the same short-term accuracy of the KNF model with LSTM networks. Nonetheless, the reduction of the statistical error during training does not follow the same trend as the instantaneous error. This suggests that using more sophisticated loss functions, including not only the instantaneous predictions but also the averaged behavior of the flow, is the simplest way to learn both aspects of the flow predictions. 
It is however remarkable that using a simple loss function based on instantaneous values we also obtained very good reproductions of Poincar\'e maps and Lyapunov exponents. We also assessed the impact of the time step, where the best network performance was obtained with $\Delta t=1$. Additionally, we compared the performance of LSTM networks and GRUs, and the former clearly provided much better predictions. 

The methods described in this work can be extended for their use in non-intrusive sensing applications~\citep{epod} or in advanced flow-control methods~\citep{rabault2}, among others. In particular, the excellent short-term prediction capabilities of the KNF method may help to increase the temporal resolution of experimental measurements~\citep{andrea_stefano}, which may aid in the assessment of the dynamics of coherent structures in turbulent flows.

\section*{Acknowledgments}

Note that all the codes employed in this work will be released as open-source on GitHub upon publication in the peer-reviewed literature. The authors acknowledge the funding provided by the Swedish e-Science Research Centre (SeRC), the G\"oran Gustafsson Foundation and the Knut and Alice Wallenberg (KAW) Foundation. Part of the analysis was performed on resources provided by the Swedish National Infrastructure for Computing (SNIC) at PDC and HPC2N.

\bibliographystyle{elsarticle-harv}
\biboptions{authoryear}
\bibliography{References}

\begin{thebibliography}{69}
\expandafter\ifx\csname natexlab\endcsname\relax\def\natexlab#1{#1}\fi
\providecommand{\url}[1]{\texttt{#1}}
\providecommand{\href}[2]{#2}
\providecommand{\path}[1]{#1}
\providecommand{\DOIprefix}{doi:}
\providecommand{\ArXivprefix}{arXiv:}
\providecommand{\URLprefix}{URL: }
\providecommand{\Pubmedprefix}{pmid:}
\providecommand{\doi}[1]{\href{http://dx.doi.org/#1}{\path{#1}}}
\providecommand{\Pubmed}[1]{\href{pmid:#1}{\path{#1}}}
\providecommand{\bibinfo}[2]{#2}
\ifx\xfnm\relax \def\xfnm[#1]{\unskip,\space#1}\fi
\bibitem[{Abadi et~al.(2016)}]{tensor_flow}
\bibinfo{author}{Abadi, M.}, et~al., \bibinfo{year}{2016}.
\newblock \bibinfo{title}{Tensorflow: a system for large-scale machine
  learning.}
\newblock \bibinfo{journal}{In Proc. 12th USENIX Symposium on Operating Systems
  Design and Implementation (OSDI ’16)} \bibinfo{volume}{16},
  \bibinfo{pages}{265--283}.
\bibitem[{Abarbanel et~al.(1993)Abarbanel, Brown, Sidorowich and
  Tsimring}]{Abarbanel1993}
\bibinfo{author}{Abarbanel, H.D.I.}, \bibinfo{author}{Brown, R.},
  \bibinfo{author}{Sidorowich, J.J.}, \bibinfo{author}{Tsimring, L.S.},
  \bibinfo{year}{1993}.
\newblock \bibinfo{title}{The analysis of observed chaotic data in physical
  systems}.
\newblock \bibinfo{journal}{Rev. Mod. Phys.} \bibinfo{volume}{65},
  \bibinfo{pages}{1331--1392}.
\bibitem[{Arbabi and Mezi\ifmmode~\acute{c}\else \'{c}\fi{}(2017)}]{Arbabi}
\bibinfo{author}{Arbabi, H.}, \bibinfo{author}{Mezi\ifmmode~\acute{c}\else
  \'{c}\fi{}, I.}, \bibinfo{year}{2017}.
\newblock \bibinfo{title}{Study of dynamics in post-transient flows using
  {Koopman} mode decomposition}.
\newblock \bibinfo{journal}{Phys. Rev. Fluids} \bibinfo{volume}{2},
  \bibinfo{pages}{124402}.
\bibitem[{Arbabi and Mezić(2017)}]{Arbabi2017}
\bibinfo{author}{Arbabi, H.}, \bibinfo{author}{Mezić, I.},
  \bibinfo{year}{2017}.
\newblock \bibinfo{title}{Ergodic theory, dynamic mode decomposition, and
  computation of spectral properties of the {Koopman} operator}.
\newblock \bibinfo{journal}{SIAM J. Appl. Dyn. Syst.} \bibinfo{volume}{16},
  \bibinfo{pages}{2096--2126}.
\bibitem[{Beck et~al.(2019)Beck, Flad and Munz}]{beck_et_al}
\bibinfo{author}{Beck, A.D.}, \bibinfo{author}{Flad, D.G.},
  \bibinfo{author}{Munz, C.D.}, \bibinfo{year}{2019}.
\newblock \bibinfo{title}{{Deep neural networks for data-driven LES closure
  models}}.
\newblock \bibinfo{journal}{J. Comput. Phys.} \bibinfo{volume}{398},
  \bibinfo{pages}{108910}.
\bibitem[{Berger et~al.(2015)Berger, Sastuba, Vogt, Jung and Amor}]{robotics}
\bibinfo{author}{Berger, E.}, \bibinfo{author}{Sastuba, M.},
  \bibinfo{author}{Vogt, D.}, \bibinfo{author}{Jung, B.},
  \bibinfo{author}{Amor, H.B.}, \bibinfo{year}{2015}.
\newblock \bibinfo{title}{Estimation of perturbations in robotic behavior using
  dynamic mode decomposition}.
\newblock \bibinfo{journal}{Adv. Robot} \bibinfo{volume}{29},
  \bibinfo{pages}{331--343}.
\bibitem[{Bor\'ee(2003)}]{epod}
\bibinfo{author}{Bor\'ee, J.}, \bibinfo{year}{2003}.
\newblock \bibinfo{title}{{Extended proper orthogonal decomposition: a tool to
  analyse correlated events in turbulent flows}}.
\newblock \bibinfo{journal}{Exp. Fluids} \bibinfo{volume}{35},
  \bibinfo{pages}{188--192}.
\bibitem[{Brunton et~al.(2016a)Brunton, Johnson, Ojemann and
  Kutz}]{neuroscience}
\bibinfo{author}{Brunton, B.W.}, \bibinfo{author}{Johnson, L.A.},
  \bibinfo{author}{Ojemann, J.G.}, \bibinfo{author}{Kutz, J.N.},
  \bibinfo{year}{2016}a.
\newblock \bibinfo{title}{Extracting spatial–temporal coherent patterns in
  large-scale neural recordings using dynamic mode decomposition}.
\newblock \bibinfo{journal}{J. Neurosci. Methods} \bibinfo{volume}{258},
  \bibinfo{pages}{1--15}.
\bibitem[{Brunton et~al.(2017)Brunton, Brunton, Proctor, Kaiser and
  Kutz}]{Brunton2017}
\bibinfo{author}{Brunton, S.L.}, \bibinfo{author}{Brunton, B.W.},
  \bibinfo{author}{Proctor, J.L.}, \bibinfo{author}{Kaiser, E.},
  \bibinfo{author}{Kutz, J.N.}, \bibinfo{year}{2017}.
\newblock \bibinfo{title}{Chaos as an intermittently forced linear system}.
\newblock \bibinfo{journal}{Nat. Commun.} \bibinfo{volume}{8},
  \bibinfo{pages}{19}.
\bibitem[{Brunton et~al.(2020)Brunton, Noack and
  Koumoutsakos}]{brunton_et_al_2020}
\bibinfo{author}{Brunton, S.L.}, \bibinfo{author}{Noack, B.R.},
  \bibinfo{author}{Koumoutsakos, P.}, \bibinfo{year}{2020}.
\newblock \bibinfo{title}{Machine learning for fluid mechanics}.
\newblock \bibinfo{journal}{Annu. Rev. Fluid Mech.} \bibinfo{volume}{52},
  \bibinfo{pages}{477--508}.
\bibitem[{Brunton et~al.(2016b)Brunton, Proctor and Kutz}]{Brunton2016}
\bibinfo{author}{Brunton, S.L.}, \bibinfo{author}{Proctor, J.L.},
  \bibinfo{author}{Kutz, J.N.}, \bibinfo{year}{2016}b.
\newblock \bibinfo{title}{Discovering governing equations from data by sparse
  identification of nonlinear dynamical systems}.
\newblock \bibinfo{journal}{PANS} \bibinfo{volume}{113},
  \bibinfo{pages}{3932--3937}.
\bibitem[{Budišić et~al.(2012)Budišić, Mohr and Mezić}]{AppliedKoopmanism}
\bibinfo{author}{Budišić, M.}, \bibinfo{author}{Mohr, R.},
  \bibinfo{author}{Mezić, I.}, \bibinfo{year}{2012}.
\newblock \bibinfo{title}{Applied {Koopmanism}}.
\newblock \bibinfo{journal}{Chaos} \bibinfo{volume}{22},
  \bibinfo{pages}{047510}.
\bibitem[{Chiappa et~al.(2017)Chiappa, Racani{\`e}re, Wierstra and
  Mohamed}]{Chiappa}
\bibinfo{author}{Chiappa, S.}, \bibinfo{author}{Racani{\`e}re, S.},
  \bibinfo{author}{Wierstra, D.}, \bibinfo{author}{Mohamed, S.},
  \bibinfo{year}{2017}.
\newblock \bibinfo{title}{Recurrent environment simulators}.
\newblock \bibinfo{journal}{In Proc. 5th International Conference on Learning
  Representations} .
\bibitem[{Cho et~al.(2014)Cho, Bahdanau, Bougares, Schwenk and Bengio}]{cho}
\bibinfo{author}{Cho, K.}, \bibinfo{author}{Bahdanau, D.},
  \bibinfo{author}{Bougares, F.}, \bibinfo{author}{Schwenk, H.},
  \bibinfo{author}{Bengio, Y.}, \bibinfo{year}{2014}.
\newblock \bibinfo{title}{Learning phrase representations using {RNN}
  {E}ncoder{--D}ecoder for statistical machine translation}, in:
  \bibinfo{booktitle}{In Proc. 2014 Conference on Empirical Methods in Natural
  Language Processing}, \bibinfo{publisher}{ACL}. pp.
  \bibinfo{pages}{1724--1734}.
\bibitem[{Crutchfield and McNamara(1987)}]{Crutchfield1987}
\bibinfo{author}{Crutchfield, J.P.}, \bibinfo{author}{McNamara, B.S.},
  \bibinfo{year}{1987}.
\newblock \bibinfo{title}{Equations of motion from a data series}.
\newblock \bibinfo{journal}{Complex Systems} \bibinfo{volume}{1},
  \bibinfo{pages}{417--452}.
\bibitem[{Cybenko(1989)}]{Cybenko}
\bibinfo{author}{Cybenko, G.}, \bibinfo{year}{1989}.
\newblock \bibinfo{title}{Approximation by superpositions of a sigmoidal
  function}.
\newblock \bibinfo{journal}{G. Math. Control Signal Systems}
  \bibinfo{volume}{2}, \bibinfo{pages}{303--314}.
\bibitem[{De~Fauw et~al.(2018)De~Fauw, Ledsam and
  Romera-Paredes}]{defauw_et_al_2018}
\bibinfo{author}{De~Fauw, J.}, \bibinfo{author}{Ledsam, J.},
  \bibinfo{author}{Romera-Paredes, B.e.a.}, \bibinfo{year}{2018}.
\newblock \bibinfo{title}{{Clinically applicable deep learning for diagnosis
  and referral in retinal disease}}.
\newblock \bibinfo{journal}{Nat. Med.} \bibinfo{volume}{24},
  \bibinfo{pages}{1342--1350}.
\bibitem[{Discetti et~al.(2019)Discetti, Bellani, \"Orl\"u, Serpieri,
  Sanmiguel~Vila, Raiola, Zheng, Mascotelli, Talamelli and
  Ianiro}]{andrea_stefano}
\bibinfo{author}{Discetti, S.}, \bibinfo{author}{Bellani, G.},
  \bibinfo{author}{\"Orl\"u, R.}, \bibinfo{author}{Serpieri, J.},
  \bibinfo{author}{Sanmiguel~Vila, C.}, \bibinfo{author}{Raiola, M.},
  \bibinfo{author}{Zheng, X.}, \bibinfo{author}{Mascotelli, L.},
  \bibinfo{author}{Talamelli, A.}, \bibinfo{author}{Ianiro, A.},
  \bibinfo{year}{2019}.
\newblock \bibinfo{title}{{Characterization of very-large-scale motions in
  high-$Re$ pipe flows}}.
\newblock \bibinfo{journal}{Exp. Therm. Fluid Sci.} \bibinfo{volume}{104},
  \bibinfo{pages}{1--8}.
\bibitem[{Duraisamy et~al.(2019)Duraisamy, Iaccarino and
  Xiao}]{duraisamy_et_al}
\bibinfo{author}{Duraisamy, K.}, \bibinfo{author}{Iaccarino, G.},
  \bibinfo{author}{Xiao, H.}, \bibinfo{year}{2019}.
\newblock \bibinfo{title}{Turbulence modeling in the age of data}.
\newblock \bibinfo{journal}{Annu. Rev. Fluid Mech.} \bibinfo{volume}{51},
  \bibinfo{pages}{357--377}.
\bibitem[{Farmer and Sidorowich(1987)}]{Farmer1987}
\bibinfo{author}{Farmer, J.D.}, \bibinfo{author}{Sidorowich, J.J.},
  \bibinfo{year}{1987}.
\newblock \bibinfo{title}{Predicting chaotic time series}.
\newblock \bibinfo{journal}{Phys. Rev. Lett.} \bibinfo{volume}{59},
  \bibinfo{pages}{845--848}.
\bibitem[{Fukami et~al.(2019a)Fukami, Fukagata and Taira}]{fukami2019}
\bibinfo{author}{Fukami, K.}, \bibinfo{author}{Fukagata, K.},
  \bibinfo{author}{Taira, K.}, \bibinfo{year}{2019}a.
\newblock \bibinfo{title}{{Super-resolution reconstruction of turbulent flows
  with machine learning}}.
\newblock \bibinfo{journal}{J. Fluid Mech.} \bibinfo{volume}{870},
  \bibinfo{pages}{106--120}.
\bibitem[{Fukami et~al.(2019b)Fukami, Nabae, Kawai and Fukagata}]{fukami_et_al}
\bibinfo{author}{Fukami, K.}, \bibinfo{author}{Nabae, Y.},
  \bibinfo{author}{Kawai, K.}, \bibinfo{author}{Fukagata, K.},
  \bibinfo{year}{2019}b.
\newblock \bibinfo{title}{Synthetic turbulent inflow generator using machine
  learning}.
\newblock \bibinfo{journal}{Phys. Rev. Fluids} \bibinfo{volume}{4},
  \bibinfo{pages}{064603}.
\bibitem[{{Gavish} and {Donoho}(2014)}]{Gavish2014}
\bibinfo{author}{{Gavish}, M.}, \bibinfo{author}{{Donoho}, D.L.},
  \bibinfo{year}{2014}.
\newblock \bibinfo{title}{The optimal hard threshold for singular values is
  $4/\sqrt {3}$}.
\newblock \bibinfo{journal}{IEEE Trans. Inf. Theory} \bibinfo{volume}{60},
  \bibinfo{pages}{5040--5053}.
\bibitem[{Giannakis et~al.(2018)Giannakis, Kolchinskaya, Krasnov and
  Schumacher}]{giannakis2018}
\bibinfo{author}{Giannakis, D.}, \bibinfo{author}{Kolchinskaya, A.},
  \bibinfo{author}{Krasnov, D.}, \bibinfo{author}{Schumacher, J.},
  \bibinfo{year}{2018}.
\newblock \bibinfo{title}{{Koopman} analysis of the long-term evolution in a
  turbulent convection cell}.
\newblock \bibinfo{journal}{J. Fluid Mech.} \bibinfo{volume}{847},
  \bibinfo{pages}{735–767}.
\bibitem[{Guastoni et~al.(2020a)Guastoni, Encinar, Schlatter, Azizpour and
  Vinuesa}]{guastoni}
\bibinfo{author}{Guastoni, L.}, \bibinfo{author}{Encinar, M.P.},
  \bibinfo{author}{Schlatter, P.}, \bibinfo{author}{Azizpour, H.},
  \bibinfo{author}{Vinuesa, R.}, \bibinfo{year}{2020}a.
\newblock \bibinfo{title}{{Prediction of wall-bounded turbulence from wall
  quantities using convolutional neural networks}}.
\newblock \bibinfo{journal}{J. Phys.: Conf. Ser.} \bibinfo{volume}{1522},
  \bibinfo{pages}{012022}.
\bibitem[{Guastoni et~al.(2020b)Guastoni, G\"uemes, Ianiro, Discetti,
  Schlatter, Azizpour and Vinuesa}]{guastoni2}
\bibinfo{author}{Guastoni, L.}, \bibinfo{author}{G\"uemes, A.},
  \bibinfo{author}{Ianiro, A.}, \bibinfo{author}{Discetti, S.},
  \bibinfo{author}{Schlatter, P.}, \bibinfo{author}{Azizpour, H.},
  \bibinfo{author}{Vinuesa, R.}, \bibinfo{year}{2020}b.
\newblock \bibinfo{title}{{Convolutional-network models to predict wall-bounded
  turbulence from wall quantities}}.
\newblock \bibinfo{journal}{arXiv:2006.12483} .
\bibitem[{Guastoni et~al.(2019)Guastoni, Srinivasan, Azizpour, Schlatter and
  Vinuesa}]{guastoni_tsfp}
\bibinfo{author}{Guastoni, L.}, \bibinfo{author}{Srinivasan, P.A.},
  \bibinfo{author}{Azizpour, H.}, \bibinfo{author}{Schlatter, P.},
  \bibinfo{author}{Vinuesa, R.}, \bibinfo{year}{2019}.
\newblock \bibinfo{title}{{On the use of recurrent neural networks for
  predictions of turbulent flows}}.
\newblock \bibinfo{journal}{Proc. Intern. Symp. on Turbulence \& Shear Flow
  Phenomena (TSFP-11), Southampton, UK, July 30 -- August 2} .
\bibitem[{G{\"u}emes et~al.(2019)G{\"u}emes, Discetti and Ianiro}]{guemes}
\bibinfo{author}{G{\"u}emes, A.}, \bibinfo{author}{Discetti, S.},
  \bibinfo{author}{Ianiro, A.}, \bibinfo{year}{2019}.
\newblock \bibinfo{title}{{Sensing the turbulent large-scale motions with their
  wall signature}}.
\newblock \bibinfo{journal}{Phys. Fluids} \bibinfo{volume}{31},
  \bibinfo{pages}{125112}.
\bibitem[{Ham et~al.(2019)Ham, Kim and Luo}]{ham_et_al_2019}
\bibinfo{author}{Ham, Y.G.}, \bibinfo{author}{Kim, J.H.}, \bibinfo{author}{Luo,
  J.J.}, \bibinfo{year}{2019}.
\newblock \bibinfo{title}{{Deep learning for multi-year ENSO forecasts}}.
\newblock \bibinfo{journal}{Nature} \bibinfo{volume}{573},
  \bibinfo{pages}{568--572}.
\bibitem[{Hochreiter and Schmidhuber(1997)}]{hochreiter_schmidhuber}
\bibinfo{author}{Hochreiter, S.}, \bibinfo{author}{Schmidhuber, J.},
  \bibinfo{year}{1997}.
\newblock \bibinfo{title}{Long short-term memory}.
\newblock \bibinfo{journal}{Neural Comput.} \bibinfo{volume}{9},
  \bibinfo{pages}{1735--1780}.
\bibitem[{Jean et~al.(2016)Jean, Burke, Xie, Davis, Lobell and
  Ermon}]{jean_et_al_2016}
\bibinfo{author}{Jean, N.}, \bibinfo{author}{Burke, M.}, \bibinfo{author}{Xie,
  M.}, \bibinfo{author}{Davis, W.M.}, \bibinfo{author}{Lobell, D.B.},
  \bibinfo{author}{Ermon, S.}, \bibinfo{year}{2016}.
\newblock \bibinfo{title}{{Combining satellite imagery and machine learning to
  predict poverty}}.
\newblock \bibinfo{journal}{Science} \bibinfo{volume}{353},
  \bibinfo{pages}{790--794}.
\bibitem[{Jim\'enez(2018)}]{jimenez_ml}
\bibinfo{author}{Jim\'enez, J.}, \bibinfo{year}{2018}.
\newblock \bibinfo{title}{Machine-aided turbulence theory}.
\newblock \bibinfo{journal}{J. Fluid Mech.} \bibinfo{volume}{854, R1},
  \bibinfo{pages}{1--11}.
\bibitem[{Khodkar et~al.(2019)Khodkar, Hassanzadeh and Antoulas}]{khodkar2019}
\bibinfo{author}{Khodkar, M.A.}, \bibinfo{author}{Hassanzadeh, P.},
  \bibinfo{author}{Antoulas, A.}, \bibinfo{year}{2019}.
\newblock \bibinfo{title}{A {Koopman}-based framework for forecasting the
  spatiotemporal evolution of chaotic dynamics with nonlinearities modeled as
  exogenous forcings}.
\newblock \bibinfo{journal}{arXiv preprint arXiv:1909.00076} .
\bibitem[{Koopman(1931)}]{Koopman}
\bibinfo{author}{Koopman, B.O.}, \bibinfo{year}{1931}.
\newblock \bibinfo{title}{Hamiltonian systems and transformation in {Hilbert}
  space}.
\newblock \bibinfo{journal}{Proc. Natl. Acad. Sci.} \bibinfo{volume}{17},
  \bibinfo{pages}{315--318}.
\bibitem[{Koopman and Neumann(1932)}]{Koopman2}
\bibinfo{author}{Koopman, B.O.}, \bibinfo{author}{Neumann, J.V.},
  \bibinfo{year}{1932}.
\newblock \bibinfo{title}{Dynamical systems of continuous spectra}.
\newblock \bibinfo{journal}{Proc. Natl. Acad. Sci.} \bibinfo{volume}{18},
  \bibinfo{pages}{255--263}.
\bibitem[{Kuramoto and Tsuzuki(1976)}]{Kuramoto}
\bibinfo{author}{Kuramoto, Y.}, \bibinfo{author}{Tsuzuki, T.},
  \bibinfo{year}{1976}.
\newblock \bibinfo{title}{{Persistent Propagation of Concentration Waves in
  Dissipative Media Far from Thermal Equilibrium}}.
\newblock \bibinfo{journal}{Progress of Theoretical Physics}
  \bibinfo{volume}{55}, \bibinfo{pages}{356--369}.
\bibitem[{Kutz(2017)}]{kutz}
\bibinfo{author}{Kutz, J.N.}, \bibinfo{year}{2017}.
\newblock \bibinfo{title}{Deep learning in fluid dynamics}.
\newblock \bibinfo{journal}{J. Fluid Mech.} \bibinfo{volume}{814},
  \bibinfo{pages}{1--4}.
\bibitem[{Kutz et~al.(2016)Kutz, Fu and Brunton}]{MultiresolutionDMD}
\bibinfo{author}{Kutz, J.N.}, \bibinfo{author}{Fu, X.},
  \bibinfo{author}{Brunton, S.L.}, \bibinfo{year}{2016}.
\newblock \bibinfo{title}{Multiresolution dynamic mode decomposition}.
\newblock \bibinfo{journal}{SIAM J. Appl. Dyn. Syst.} \bibinfo{volume}{15},
  \bibinfo{pages}{713--735}.
\bibitem[{Lapeyre et~al.(2019)Lapeyre, Misdariis, Cazard, Veynante and
  Poinsot}]{lapeyre_et_al}
\bibinfo{author}{Lapeyre, C.J.}, \bibinfo{author}{Misdariis, A.},
  \bibinfo{author}{Cazard, N.}, \bibinfo{author}{Veynante, D.},
  \bibinfo{author}{Poinsot, T.}, \bibinfo{year}{2019}.
\newblock \bibinfo{title}{Training convolutional neural networks to estimate
  turbulent sub-grid scale reaction rates}.
\newblock \bibinfo{journal}{Combust. Flame} \bibinfo{volume}{203},
  \bibinfo{pages}{255}.
\bibitem[{Li et~al.(2017)Li, Dietrich, Bollt and Kevrekidis}]{Li2017}
\bibinfo{author}{Li, Q.}, \bibinfo{author}{Dietrich, F.},
  \bibinfo{author}{Bollt, E.M.}, \bibinfo{author}{Kevrekidis, I.G.},
  \bibinfo{year}{2017}.
\newblock \bibinfo{title}{Extended dynamic mode decomposition with dictionary
  learning: A data-driven adaptive spectral decomposition of the {Koopman}
  operator}.
\newblock \bibinfo{journal}{Chaos} \bibinfo{volume}{27},
  \bibinfo{pages}{103111}.
\bibitem[{Ling et~al.(2016)Ling, Kurzawski and Templeton}]{ling_et_al}
\bibinfo{author}{Ling, J.}, \bibinfo{author}{Kurzawski, A.},
  \bibinfo{author}{Templeton, J.}, \bibinfo{year}{2016}.
\newblock \bibinfo{title}{Reynolds averaged turbulence modelling using deep
  neural networks with embedded invariance}.
\newblock \bibinfo{journal}{J. Fluid Mech.} \bibinfo{volume}{807},
  \bibinfo{pages}{155--166}.
\bibitem[{Lorenz(2006)}]{lorenz96}
\bibinfo{author}{Lorenz, E.N.}, \bibinfo{year}{2006}.
\newblock \bibinfo{title}{Predictability–a problem partly solved}.
  \bibinfo{publisher}{Cambridge University Press}.
\newblock p. \bibinfo{pages}{40–58}.
\bibitem[{Lusch et~al.(2018)Lusch, Kutz and Brunton}]{Lusch2018}
\bibinfo{author}{Lusch, B.}, \bibinfo{author}{Kutz, J.N.},
  \bibinfo{author}{Brunton, S.L.}, \bibinfo{year}{2018}.
\newblock \bibinfo{title}{Deep learning for universal linear embeddings of
  nonlinear dynamics}.
\newblock \bibinfo{journal}{Nat. Commun.} \bibinfo{volume}{9},
  \bibinfo{pages}{4950}.
\bibitem[{{Mauroy} and {Goncalves}(2016)}]{Mauroy}
\bibinfo{author}{{Mauroy}, A.}, \bibinfo{author}{{Goncalves}, J.},
  \bibinfo{year}{2016}.
\newblock \bibinfo{title}{Linear identification of nonlinear systems: A lifting
  technique based on the {Koopman} operator}, in: \bibinfo{booktitle}{2016 IEEE
  55th Conference on Decision and Control (CDC)}, pp.
  \bibinfo{pages}{6500--6505}.
\bibitem[{Mezić(2005)}]{Mezic2005}
\bibinfo{author}{Mezić, I.}, \bibinfo{year}{2005}.
\newblock \bibinfo{title}{Spectral properties of dynamical systems, model
  reduction and decompositions}.
\newblock \bibinfo{journal}{Nonlinear Dyn.} \bibinfo{volume}{41},
  \bibinfo{pages}{309--325}.
\bibitem[{Mezić(2013)}]{Mezic2013}
\bibinfo{author}{Mezić, I.}, \bibinfo{year}{2013}.
\newblock \bibinfo{title}{Analysis of fluid flows via spectral properties of
  the {Koopman} operator}.
\newblock \bibinfo{journal}{Annu. Rev. Fluid Mech.} \bibinfo{volume}{45},
  \bibinfo{pages}{357--378}.
\bibitem[{Mezić and Banaszuk(2004)}]{Mezic2004}
\bibinfo{author}{Mezić, I.}, \bibinfo{author}{Banaszuk, A.},
  \bibinfo{year}{2004}.
\newblock \bibinfo{title}{Comparison of systems with complex behavior}.
\newblock \bibinfo{journal}{Physica D} \bibinfo{volume}{197},
  \bibinfo{pages}{101--133}.
\bibitem[{Milano and Koumoutsakos(2002)}]{milano_koumoutsakos}
\bibinfo{author}{Milano, M.}, \bibinfo{author}{Koumoutsakos, P.},
  \bibinfo{year}{2002}.
\newblock \bibinfo{title}{Neural network modeling for near wall turbulent
  flow}.
\newblock \bibinfo{journal}{J. Comput. Phys.} \bibinfo{volume}{182},
  \bibinfo{pages}{1--26}.
\bibitem[{Moehlis et~al.(2004)Moehlis, Faisst and Eckhardt}]{moehlis_et_al}
\bibinfo{author}{Moehlis, J.}, \bibinfo{author}{Faisst, H.},
  \bibinfo{author}{Eckhardt, B.}, \bibinfo{year}{2004}.
\newblock \bibinfo{title}{A low-dimensional model for turbulent shear flows}.
\newblock \bibinfo{journal}{New J. Phys.} \bibinfo{volume}{6},
  \bibinfo{pages}{56}.
\bibitem[{Norouzzadeh et~al.(2018)Norouzzadeh, Nguyen, Kosmala, Swanson,
  Palmer, Packer and Clune}]{norouzzadeh_et_al_2018}
\bibinfo{author}{Norouzzadeh, M.S.}, \bibinfo{author}{Nguyen, A.},
  \bibinfo{author}{Kosmala, M.}, \bibinfo{author}{Swanson, A.},
  \bibinfo{author}{Palmer, M.S.}, \bibinfo{author}{Packer, C.},
  \bibinfo{author}{Clune, J.}, \bibinfo{year}{2018}.
\newblock \bibinfo{title}{{Automatically identifying, counting, and describing
  wild animals in camera-trap images with deep learning}}.
\newblock \bibinfo{journal}{Proc. Natl Acad. Sci.} \bibinfo{volume}{115},
  \bibinfo{pages}{E5716--E5725}.
\bibitem[{Page and Kerswell(2019)}]{page2019}
\bibinfo{author}{Page, J.}, \bibinfo{author}{Kerswell, R.R.},
  \bibinfo{year}{2019}.
\newblock \bibinfo{title}{{Koopman} mode expansions between simple invariant
  solutions}.
\newblock \bibinfo{journal}{J. Fluid Mech.} \bibinfo{volume}{879},
  \bibinfo{pages}{1–27}.
\bibitem[{Pandey et~al.(2020)Pandey, Schumacher and Sreenivasan}]{Pandey_esn}
\bibinfo{author}{Pandey, S.}, \bibinfo{author}{Schumacher, J.},
  \bibinfo{author}{Sreenivasan, K.R.}, \bibinfo{year}{2020}.
\newblock \bibinfo{title}{A perspective on machine learning in turbulent
  flows}.
\newblock \bibinfo{journal}{J. Turbul.} \bibinfo{volume}{0},
  \bibinfo{pages}{1--18}.
\bibitem[{Proctor et~al.(2016)Proctor, Brunton and Kutz}]{Proctor2016}
\bibinfo{author}{Proctor, J.L.}, \bibinfo{author}{Brunton, S.L.},
  \bibinfo{author}{Kutz, J.N.}, \bibinfo{year}{2016}.
\newblock \bibinfo{title}{Dynamic mode decomposition with control}.
\newblock \bibinfo{journal}{SIAM J. Appl. Dyn. Syst.} \bibinfo{volume}{15},
  \bibinfo{pages}{142--161}.
\bibitem[{Rabault et~al.(2019)Rabault, Kuchta, Jensen, Reglade and
  Cerardi}]{rabault}
\bibinfo{author}{Rabault, J.}, \bibinfo{author}{Kuchta, M.},
  \bibinfo{author}{Jensen, A.}, \bibinfo{author}{Reglade, U.},
  \bibinfo{author}{Cerardi, N.}, \bibinfo{year}{2019}.
\newblock \bibinfo{title}{{Artificial neural networks trained through deep
  reinforcement learning discover control strategies for active flow control}}.
\newblock \bibinfo{journal}{J. Fluid Mech.} \bibinfo{volume}{865},
  \bibinfo{pages}{281--302}.
\bibitem[{Raissi et~al.(2020)Raissi, Yazdani and Karniadakis}]{raissi_et_al}
\bibinfo{author}{Raissi, M.}, \bibinfo{author}{Yazdani, A.},
  \bibinfo{author}{Karniadakis, G.E.}, \bibinfo{year}{2020}.
\newblock \bibinfo{title}{{Hidden fluid mechanics: Learning velocity and
  pressure fields from flow visualizations}}.
\newblock \bibinfo{journal}{Science} \bibinfo{volume}{367},
  \bibinfo{pages}{1026--1030}.
\bibitem[{Rowley et~al.(2009)Rowley, Mezi{\'c}, Bagheri, Schlatter and
  Henningson}]{rowley}
\bibinfo{author}{Rowley, C.W.}, \bibinfo{author}{Mezi{\'c}, I.},
  \bibinfo{author}{Bagheri, S.}, \bibinfo{author}{Schlatter, P.},
  \bibinfo{author}{Henningson, D.S.}, \bibinfo{year}{2009}.
\newblock \bibinfo{title}{Spectral analysis of nonlinear flows}.
\newblock \bibinfo{journal}{J. Fluid Mech.} \bibinfo{volume}{641},
  \bibinfo{pages}{115–127}.
\bibitem[{Rumelhart et~al.(1985)Rumelhart, Hinton and
  Williams}]{rumelhart1985learning}
\bibinfo{author}{Rumelhart, D.E.}, \bibinfo{author}{Hinton, G.E.},
  \bibinfo{author}{Williams, R.J.}, \bibinfo{year}{1985}.
\newblock \bibinfo{title}{Learning internal representations by error
  propagation}.
\newblock \bibinfo{type}{Technical Report}. California Univ San Diego La Jolla
  Inst for Cognitive Science.
\bibitem[{Schmid(2010)}]{Schmid2010}
\bibinfo{author}{Schmid, P.J.}, \bibinfo{year}{2010}.
\newblock \bibinfo{title}{{Dynamic mode decomposition of numerical and
  experimental data}}.
\newblock \bibinfo{journal}{J. Fluid Mech.} \bibinfo{volume}{656},
  \bibinfo{pages}{5--28}.
\bibitem[{Sivashinsky(1982)}]{Sivashinsky}
\bibinfo{author}{Sivashinsky, G.}, \bibinfo{year}{1982}.
\newblock \bibinfo{title}{Large cells in nonlinear marangoni convection}.
\newblock \bibinfo{journal}{Physica D} \bibinfo{volume}{4}, \bibinfo{pages}{227
  -- 235}.
\bibitem[{Srinivasan et~al.(2019)Srinivasan, Guastoni, Azizpour, Schlatter and
  Vinuesa}]{srinivasan}
\bibinfo{author}{Srinivasan, P.A.}, \bibinfo{author}{Guastoni, L.},
  \bibinfo{author}{Azizpour, H.}, \bibinfo{author}{Schlatter, P.},
  \bibinfo{author}{Vinuesa, R.}, \bibinfo{year}{2019}.
\newblock \bibinfo{title}{Predictions of turbulent shear flows using deep
  neural networks}.
\newblock \bibinfo{journal}{Phys. Rev. Fluids} \bibinfo{volume}{4},
  \bibinfo{pages}{054603}.
\bibitem[{Sugihara et~al.(2012)Sugihara, May, Ye, Hsieh, Deyle, Fogarty and
  Munch}]{Sugihara496}
\bibinfo{author}{Sugihara, G.}, \bibinfo{author}{May, R.}, \bibinfo{author}{Ye,
  H.}, \bibinfo{author}{Hsieh, C.h.}, \bibinfo{author}{Deyle, E.},
  \bibinfo{author}{Fogarty, M.}, \bibinfo{author}{Munch, S.},
  \bibinfo{year}{2012}.
\newblock \bibinfo{title}{Detecting causality in complex ecosystems}.
\newblock \bibinfo{journal}{Science} \bibinfo{volume}{338},
  \bibinfo{pages}{496--500}.
\bibitem[{Takeishi et~al.(2017)Takeishi, Kawahara and Yairi}]{Takeishi2017}
\bibinfo{author}{Takeishi, N.}, \bibinfo{author}{Kawahara, Y.},
  \bibinfo{author}{Yairi, T.}, \bibinfo{year}{2017}.
\newblock \bibinfo{title}{Learning {Koopman} invariant subspaces for dynamic
  mode decomposition}, in: \bibinfo{booktitle}{NIPS 30}, pp.
  \bibinfo{pages}{1130--1140}.
\bibitem[{Takens(1981)}]{Takens}
\bibinfo{author}{Takens, F.}, \bibinfo{year}{1981}.
\newblock \bibinfo{title}{Detecting strange attractors in turbulence}, in:
  \bibinfo{booktitle}{Dynamical Systems and Turbulence, Warwick 1980},
  \bibinfo{publisher}{Springer Berlin Heidelberg}, \bibinfo{address}{Berlin,
  Heidelberg}. pp. \bibinfo{pages}{366--381}.
\bibitem[{Tang et~al.(2020)Tang, Rabault, Kuhnle, Wang and Wang}]{rabault2}
\bibinfo{author}{Tang, H.}, \bibinfo{author}{Rabault, J.},
  \bibinfo{author}{Kuhnle, A.}, \bibinfo{author}{Wang, Y.},
  \bibinfo{author}{Wang, T.}, \bibinfo{year}{2020}.
\newblock \bibinfo{title}{{Robust active flow control over a range of Reynolds
  numbers using an artificial neural network trained through deep reinforcement
  learning}}.
\newblock \bibinfo{journal}{Preprint arXiv:2004.12417} .
\bibitem[{Tu et~al.(2014)Tu, Rowley, Luchtenburg, Brunton and Kutz}]{Tu2014}
\bibinfo{author}{Tu, J.H.}, \bibinfo{author}{Rowley, C.W.},
  \bibinfo{author}{Luchtenburg, D.M.}, \bibinfo{author}{Brunton, S.L.},
  \bibinfo{author}{Kutz, J.N.}, \bibinfo{year}{2014}.
\newblock \bibinfo{title}{On dynamic mode decomposition: Theory and
  applications}.
\newblock \bibinfo{journal}{J. Comput. Dyn.} \bibinfo{volume}{1},
  \bibinfo{pages}{391--421}.
\bibitem[{Udrescu and Tegmark(2020)}]{udrescu}
\bibinfo{author}{Udrescu, S.M.}, \bibinfo{author}{Tegmark, M.},
  \bibinfo{year}{2020}.
\newblock \bibinfo{title}{{AI Feynman: A physics-inspired method for symbolic
  regression}}.
\newblock \bibinfo{journal}{Sci. Adv.} \bibinfo{volume}{6},
  \bibinfo{pages}{1--16}.
\bibitem[{Vinuesa et~al.(2020)Vinuesa, Azizpour, Leite, Balaam, Dignum,
  Domisch, Fell{\"a}nder, Langhans, Tegmark and
  Fuso~Nerini}]{vinuesa_et_al_2020}
\bibinfo{author}{Vinuesa, R.}, \bibinfo{author}{Azizpour, H.},
  \bibinfo{author}{Leite, I.}, \bibinfo{author}{Balaam, M.},
  \bibinfo{author}{Dignum, V.}, \bibinfo{author}{Domisch, S.},
  \bibinfo{author}{Fell{\"a}nder, A.}, \bibinfo{author}{Langhans, S.D.},
  \bibinfo{author}{Tegmark, M.}, \bibinfo{author}{Fuso~Nerini, F.},
  \bibinfo{year}{2020}.
\newblock \bibinfo{title}{{The role of artificial intelligence in achieving the
  Sustainable Development Goals}}.
\newblock \bibinfo{journal}{Nat. Commun.} \bibinfo{volume}{11},
  \bibinfo{pages}{233}.
\bibitem[{Williams et~al.(2015)Williams, Kevrekidis and Rowley}]{Williams2015}
\bibinfo{author}{Williams, M.O.}, \bibinfo{author}{Kevrekidis, I.G.},
  \bibinfo{author}{Rowley, C.W.}, \bibinfo{year}{2015}.
\newblock \bibinfo{title}{A data-driven approximation of the {Koopman}
  operator: Extending dynamic mode decomposition}.
\newblock \bibinfo{journal}{J Nonlinear Sci} \bibinfo{volume}{25},
  \bibinfo{pages}{1307--1346}.
\bibitem[{Wu et~al.(2018)Wu, Xiao and Paterson}]{wu_et_al}
\bibinfo{author}{Wu, J.L.}, \bibinfo{author}{Xiao, H.},
  \bibinfo{author}{Paterson, E.}, \bibinfo{year}{2018}.
\newblock \bibinfo{title}{{Physics-informed machine learning approach for
  augmenting turbulence models: A comprehensive framework}}.
\newblock \bibinfo{journal}{Phys. Rev. Fluids} \bibinfo{volume}{3},
  \bibinfo{pages}{074602}.

\end{thebibliography}

\end{document}